\documentclass[usenatbib]{mnras}
\usepackage{graphicx}
\usepackage{amsmath}
\usepackage{breqn}
\usepackage{amssymb}
\usepackage{hyperref}
\usepackage{xcolor}

\newcommand{\seedmass}{M_{\mathrm{seed}}}

\title[Local signatures of black hole seeds]{Signatures of black hole seeding in the local Universe: Predictions from the \texttt{BRAHMA} cosmological simulations}
\author[Bhowmick et al.]{Aklant K. Bhowmick$^{1}$,
Laura Blecha$^{2}$,
Paul Torrey$^{1}$, Rachel S Somerville$^{8}$ \newauthor
Luke Zoltan Kelley$^{4}$, Rainer Weinberger$^{6}$, 
Mark Vogelsberger$^{5}$,
Lars Hernquist$^{7}$, \newauthor
Priyamvada Natarajan$^{9,10,11}$, Jonathan Kho$^{1}$, Tiziana Di Matteo$^{3}$
\\
$^{1}$ Department of Astronomy, University of Virginia, Charlottesville, VA 22904, USA\\
$^{2}$ Dept. of Physics, University of Florida, Gainesville, FL 32611, USA \\
$^{3}$ McWilliams Center for Cosmology, 
Carnegie Mellon  University, Pittsburgh, PA 15213, USA \\
$^{4}$Department of Astronomy, University of California at Berkeley, Berkeley, CA 94720, USA\\
$^{5}$Dept. of Physics, Kavli Institute for Astrophysics and Space Research, Massachusetts Institute of Technology,
Cambridge, MA 02139, USA \\
$^{6}$Leibniz Institute for Astrophysics Potsdam (AIP), An der Sternwarte 16, 14482 Potsdam, Germany\\
$^{7}$Harvard-Smithsonian Center for Astrophysics, 60 Garden Street, Cambridge, MA 02138, USA\\
$^{8}$Center for Computational Astrophysics, Flatiron institute, New York, NY 10010, USA\\
$^{9}$Dept. of Physics, Yale University, New Haven, CT 06520,USA\\
$^{10}$Department of Astronomy, Yale University, 266 Whitney Avenue, New Haven, CT 06511, USA\\
$^{11}$Black Hole Initiative, Harvard University, 20 Garden Street, Cambridge, MA 02138, USA
}
\begin{document}
\maketitle
\begin{abstract}
The first ``seeds" of supermassive black holes~(BHs) continue to be an outstanding puzzle, and it is currently unclear whether  the imprints of early seed formation survive today. Here we examine the  signatures of seeding in the local Universe using five $[18~\mathrm{Mpc}]^3$ \texttt{BRAHMA} simulation boxes run to  $z=0$. They initialize $1.5\times10^5~M_{\odot}$ BHs using different seeding models. The first four boxes initialize BHs as heavy seeds using criteria that depend on dense \& metal-poor gas, Lyman-Werner radiation, gas spin, and environmental richness. The fifth box initializes BHs as descendants of lower mass seeds~($\sim10^3~M_{\odot}$) using a new stochastic seed model built in our previous work. We find that strong signatures of seeding survive in $\sim10^5-10^6~M_{\odot}$ local BHs hosted in $M_*\lesssim10^{9}~M_{\odot}$ dwarf galaxies. The signatures survive due to two reasons: 1) there is a substantial population of local $\sim10^5~M_{\odot}$ BHs that are ungrown relics of early seeds from $z\sim5-10$; 2) BH growth up to $\sim10^6~M_{\odot}$ is dominated by mergers all the way down to $z\sim0$. As the contribution from gas accretion increases, the signatures of seeding start to weaken in more massive $\gtrsim10^6~M_{\odot}$ BHs, and they eventually disappear for $\gtrsim10^7~M_{\odot}$ BHs. This is in contrast to high-z~($z\gtrsim5$) BH populations wherein the BH growth is fully merger dominated, which causes the seeding signatures to persist at least up to $\sim10^8~M_{\odot}$. 
The different seed models predict abundances of local $\sim10^6~M_{\odot}$ BHs ranging from $\sim0.01-0.05~\mathrm{Mpc}^{-3}$ with occupation fractions of $\sim20-100\%$ in $M_*\sim10^{9}~M_{\odot}$ galaxies. Our results highlight the potential for local $\sim10^5-10^6~M_{\odot}$ BH populations in dwarf galaxies to serve as a promising probe for BH seeding models. 

\end{abstract}


\section{Introduction}
\label{introduction}
The origins of supermassive black holes~(SMBHs with masses $\gtrsim10^6~M_{\odot}$) continues to be a key open question in our current understanding of galaxy formation. There are several promising candidates to account for them that span a wide range of initial seed masses. At the lowest mass end, the remnants of the first generation Population III stars, often referred to as ``light seeds" or ``Pop III seeds"~\citep{2001ApJ...550..372F,2001ApJ...551L..27M,2013ApJ...773...83X,2018MNRAS.480.3762S} are viable candidates with typical masses of $\sim100~M_{\odot}$. Runaway collisions of stars or black holes~(BHs) in dense nuclear star clusters~(NSCs), or the amplification of light seeds in the NSCs can form seeds as massive as $\sim10^3- 10^4~M_{\odot}~$\citep{2011ApJ...740L..42D,2014MNRAS.442.3616L,2020MNRAS.498.5652K,2021MNRAS.503.1051D,2021MNRAS.tmp.1381D,2021MNRAS.501.1413N}, referred to as ``NSC" seeds. More massive seeds can form under special conditions when the collapsing clouds of gas can circumvent the standard stellar evolution phases and directly collapse to form BHs in the very early Universe. The resulting seeds are referred to as direct collapse black holes or ``DCBH" seeds. They are typically considered to range between $\sim10^4-10^5~M_{\odot}$ ~\citep{2003ApJ...596...34B,2006MNRAS.370..289B,2006MNRAS.371.1813L,2007MNRAS.377L..64L,2018MNRAS.476.3523L,2019Natur.566...85W,2020MNRAS.492.4917L,2023MNRAS.526L..94B} but could form later on at $z \sim 6$ and also be as massive as $\sim10^8~M_{\odot}$~\citep{2024ApJ...961...76M}.  

A common feature amongst Pop III, NSC and DCBH seeds is that the vast majority of them are expected to form within dense and low metallicity gas in the early Universe. At later times, seed formation is expected to slow down due to increased metal enrichment. Therefore, we expect the signatures of seeding to be most strongly retained within the early BH populations prior to efficient metal mixing. The first BH populations discovered at high z~($z\gtrsim6$) prior to the ongoing James Webb Space Telescope~(JWST), were luminous quasars~(bolometric luminosities $\gtrsim10^{46} \rm erg~s^{-1}$) powered by BHs with masses ranging from $\sim10^9-10^{10}~M_{\odot}$~\citep{2001AJ....122.2833F,2010AJ....139..906W,2011Natur.474..616M,2015MNRAS.453.2259V,2016ApJ...833..222J,2016Banados,2017MNRAS.468.4702R,2018ApJS..237....5M,2018ApJ...869L...9W,2018Natur.553..473B,2019ApJ...872L...2M,2019AJ....157..236Y,2021ApJ...907L...1W}. Explaining the assembly of these BHs within a billion years after the Big Bang continues to be challenging for BH seeding and growth models. This is because a light $\sim10^2~M_{\odot}$ seed would require sustained super Eddington growth to reach the masses inferred for the bright, luminous $z\sim 6$ quasars, and a heavy $\sim10^5~M_{\odot}$ seed would require sustained Eddington growth. 

The advent of JWST has further pushed the observational frontier permitting detection of higher-z active galactic nuclei~(AGN) populations at lower luminosities by detecting a large population of fainter broad line~(BL) AGN candidates at $z\sim4-11$~\citep{2023ApJ...942L..17O,2023ApJ...959...39H,2023ApJ...954L...4K,2023arXiv230801230M,2023ApJ...953L..29L,2023arXiv230905714G,2024arXiv240403576K,2024A&A...685A..25A,2024arXiv240610341A}. Remarkably, several of the spectroscopically confirmed AGNs report BH mass measurements~($\sim10^6-10^8~M_\odot$) that appear to be $\sim10-100$ times more massive for their host galaxy stellar masses, compared to expectations from local galaxy scaling relations~\citep{2023ApJ...957L...7K,2024NatAs...8..126B,2024ApJ...960L...1N,2024ApJ...968...38K,2024arXiv240403576K,2024arXiv240610329D}. More recently, similarly overmassive BHs have also been observed at cosmic noon~\citep{2024arXiv240405793M}. While current BH mass and host galaxy stellar mass measurements do have substantial uncertainties~($\sim1~\rm dex$), these overmassive BHs can be used to place further constraints on BH seeding and growth. Such exceptionally high $M_{bh}/M_*$ ratios were predicted as a signature of heavy seeds~\citep{2013MNRAS.432.3438A,2017ApJ...838..117N,2024ApJ...960L...1N}, or efficient early growth of lower mass seeds via super Eddington accretion~(Trinca et al. in prep.) or BH-BH mergers~\citep{2024MNRAS.531.4311B}.      

By the time we get to the local Universe, the signatures of seeding are expected to be washed out by subsequent the complex mass assembly history of BHs. However, if a small fraction of seeds do form at late times as noted by \cite{2021MNRAS.501.1413N}, or if there are seeds that formed earlier but did not undergo significant growth, these populations could retain some information about their initial seeding conditions. Moreover, since the initial seed populations are expected to lie at the lowest mass end of the BH mass functions and the faintest end of the AGN luminosity functions, they would be much more readily detectable~(albeit still challenging) in the local Universe compared to higher redshifts. Detection of intermediate mass black holes or IMBHs~($M_{bh}\sim10^3-10^5~M_{\odot}$) in local dwarf galaxies~(stellar masses $M_*\lesssim10^9~M_{\odot}$) is considered to be particularly promising for learning about BH seeds~(see review by \citealt{2022NatAs...6...26R}). Over the past two decades, numerous $M_{bh}\sim10^5-10^6~M_{\odot}$ 
IMBH candidates have been detected in local dwarf galaxies~\citep{2003ApJ...588L..13F,2004ApJ...610..722G,2004ApJ...607...90B,2007ApJ...670...92G,2012ApJ...755..167D,2013ApJ...775..116R,2018MNRAS.478.2576M,2019ApJ...887..245S,2021ApJ...922..155M,2022ApJ...941...43Y,2022ApJ...936..104W,2022NatAs...6.1452A,2024arXiv240511750Z,2024SCPMA..6709811L,2024Natur.631..285H}, with the smallest detected IMBH being $\sim50000~M_{\odot}$~\citep{2015ApJ...809L..14B}. These developments have made it possible to probe the BH-galaxy scaling relations~\citep{2019ApJ...887..245S,2020ApJ...898L...3B,2021MNRAS.502L...1K,2022NatAs...6.1452A}  and BH occupation fractions~\citep{2012NatCo...3.1304G,2015ApJ...799...98M,2020ARA&A..58..257G} in the dwarf galaxy regime. Although uncertainties are substantial due to small sample sizes and challenges in BH mass measurements, they are a promising path for obtaining constraints on BH seeding and growth scenarios. 

At the theoretical end, several tools have been used to trace and make predictions for BH populations across cosmic time. Semi-analytic models~(SAMs), due to their low computational expense, are able to robustly explore a wide range of BH seeding and growth scenarios and study their impact on observed properties ~\citep{2012MNRAS.423.2533B,2018MNRAS.481.3278R,2019MNRAS.486.2336D,2021MNRAS.506..613S,2022MNRAS.511..616T,2023MNRAS.519.4753T,2023ApJ...946...51C,2023MNRAS.518.4672S}. However, SAMs do not track spatial information, like the detailed hydrodynamics of gas, which obviously plays a key role in BH formation and growth. This can only be done in full cosmological hydrodynamic simulations~\citep{2012ApJ...745L..29D,2014Natur.509..177V,2015MNRAS.452..575S,2015MNRAS.450.1349K,2015MNRAS.446..521S,2016MNRAS.460.2979V,2016MNRAS.463.3948D,2017MNRAS.467.4739K,2017MNRAS.470.1121T,2019ComAC...6....2N,2020MNRAS.498.2219V,2020NatRP...2...42V}. More recently, a class of hybrid SAMs have emerged that can extract the information about detailed gas properties from pre-existing cosmological simulations, and use them to build seed models~\citep{2020MNRAS.491.4973D,2023arXiv230911324E}. These hybrid SAMs retain the flexibility and efficiency of standard SAMs, while adding more physical realism in their seeding prescriptions. However, they are still not able to capture the impact of the BHs on the gas dynamics, since they only seed BHs in post-processing. Nevertheless, the limitations of SAM-based approaches are outshined by their computational efficiency. This has enabled them to systematically explore the impact of a wide range of seeding and growth prescriptions on BH populations all the way to the local universe~\citep{2018MNRAS.481.3278R,2023ApJ...946...51C,2023arXiv230911324E}. All these studies show that BHs within local dwarf galaxies are strongly sensitive to the modeling of both BH seeding and BH growth.

Despite having self-consistent gas hydrodynamics, the modeling of BH seeding is extremely challenging in cosmological simulations. The substantial computational expense prohibits their ability to resolve the seeds and their formation processes in a cosmic volume. With a typical gas mass resolution of $\sim10^5-10^6~M_{\odot}$, many galaxy formation simulations simply seed $\sim10^5-10^6~M_{\odot}$ BHs based on a halo or stellar mass threshold just to ensure that sufficiently massive galaxies contain SMBHs~\citep[e.g.][]{2012ApJ...745L..29D,2014Natur.509..177V,2015MNRAS.450.1349K,2019MNRAS.490.3234N,2019ComAC...6....2N}. Even though these simulations reproduce many of the observed local galaxy and BH properties~\citep{2020MNRAS.493..899H,2021MNRAS.503.1940H}, their implemented seeding prescriptions are too simplistic to provide any information about the origin of BH seeds. More recent simulations have adopted seeding models based on local gas properties that can better emulate the physical gas conditions for Pop III, NSC or DCBH seeding, for e.g., seeding BHs within high density and low metallicity gas~\citep{2016MNRAS.463..529H,2017MNRAS.467.4739K,2017MNRAS.470.1121T,2024arXiv240218773J}. 

However, the formation conditions, initial masses, and formation rates of Pop III, NSC and DCBH seeds are still very uncertain, which prevents us from ascertaining the best approaches to model them in cosmological simulations~(as well as SAMs). For example, the initial mass function of Pop III stars is unknown, but that determines the formation rates and initial masses of Pop III seeds. For NSC seeds, there are uncertainties in the NSC occupation fractions within galaxies, as well as the final BH seed masses that can be grown in an NSC via rapid amplification or runaway stellar and BH collisions~(including the impact of gravitational recoil). DCBHs may require additional conditions beyond having a  dense and pristine gas, such as strong Lyman Werner~(LW) radiation~\citep{2010MNRAS.402.1249S,2014MNRAS.445..544S,2017MNRAS.469.3329W}, low gas spins~\citep{2006MNRAS.371.1813L}, or dynamical heating induced by major mergers~\citep{2019Natur.566...85W,2020OJAp....3E..15R,2020MNRAS.492.3021R,2024arXiv241006141P}~(see also  \citealt{2021MNRAS.501.1413N,2024arXiv240517975R}). Lastly, there are also uncertainties in the modeling of BH accretion and dynamics in cosmological simulations~(as well as SAMs) due to resolution limitations. This is expected to pose yet another challenge in constraining BH seeding and growth mechanisms from observations, due to potential degeneracies between the signatures of seeding, growth by accretion and merger dynamics for BH populations. In the regime of local dwarf galaxies, \cite{2022MNRAS.514.4912H} performed one of the largest studies to date comparing BH populations within five different cosmological simulations. They found that the different cosmological simulations result in widely different BH populations within dwarf galaxies. However, the reasons for these differences can be difficult to pin-point due to the variety of treatments deployed for initial BH seeding, accretion, dynamics, as well as AGN and stellar feedback processes.  

In light of the foregoing challenges, the \texttt{BRAHMA} simulation project~\citep{2024MNRAS.531.4311B,2024MNRAS.533.1907B} was started with the aim of performing a large systematic study to quantify the impact of seeding on BH populations across cosmic time. The \texttt{BRAHMA} simulations employ a set of novel seeding prescriptions that are based on a wide range of gas properties of halos, such as dense and metal-poor gas mass, LW radiation, gas spin and halo environmental richness. The central idea is to explore different seeding model variations that can not only encompass the different seeding scenarios~(Pop III, NSC and DCBH), but also encapsulate the different physics uncertainties within each model as noted above. All these seed models were extensively tested using high resolution zoom and constrained simulations~\citep{2021MNRAS.507.2012B,2022MNRAS.510..177B,2022MNRAS.516..138B}. Furthermore, for large volume-low resolution simulations, we also built a novel stochastic seed model that can represent seeds that are $\sim10-100$ times below the gas mass resolution limit~\citep{2024MNRAS.529.3768B}. In our initial \texttt{BRAHMA} papers, we introduced a set of uniform simulation boxes~($9,18~\&~36~\mathrm{Mpc}$ box-size) and focused on making predictions for high-z BH populations for a set of low mass seed models emulating lower mass~(Pop III and NSC) seeds~\citep{2024MNRAS.531.4311B} and  heavy~(DCBH) seeds~\citep{2024MNRAS.533.1907B}. 

In this work, we use a subset of boxes from \cite{2024MNRAS.531.4311B,2024MNRAS.533.1907B} to explore signatures of seeding in the local Universe, particularly in the regime of dwarf galaxies~($M_*\lesssim10^9~M_{\odot}$). We compare our predictions with existing observational constraints for BH populations in dwarf galaxies, as well as other theoretical predictions from SAMs and hydrodynamic simulations. One advantage we have compared to \cite{2022MNRAS.514.4912H} is that apart from BH seeding, every other aspect of our galaxy formation model~(including BH accretion and dynamics) is identical; this enables us to isolate the impact of seed formation in these populations. In Section \ref{The BRAHMA simulations sec}, we summarize the key features of the \texttt{BRAHMA} simulations, along with the different seed models applied to the different boxes. Section \ref{Results sec} shows key predictions of a wide range of quantities including BH mass functions, AGN luminosity functions, $M_*-M_{bh}$ relations, and the BH occupation fractions. In Section \ref{Discussion sec}, we discuss the implications of our results on the viability of low mass (Pop III and NSC) seeds vs. heavy (DCBH) seeds as possible origins of the observed local BH populations. Lastly, Section \ref{Conclusions sec} summarizes the main conclusions of the paper.   

\section{The BRAHMA simulations}
\label{The BRAHMA simulations sec}
\label{methods}
\label{AREPO cosmological code and the IllustrisTNG model}
The \texttt{BRAHMA} simulation suite~\citep{2024MNRAS.529.3768B,2024MNRAS.533.1907B} was run using the \texttt{AREPO} gravity + magneto-hydrodynamics~(MHD) code~\citep{2010MNRAS.401..791S,2011MNRAS.418.1392P,2016MNRAS.462.2603P,2020ApJS..248...32W}. The N-body gravity solver uses the PM Tree~\citep{1986Natur.324..446B} method whereas the MHD solver uses a dynamic unstructured grid generated via a Voronoi tessellation of the domain. The initial conditions were generated using \texttt{MUSIC}~\citep{2011MNRAS.415.2101H} and adopted a  \cite{2016A&A...594A..13P} cosmology i.e. $\Omega_{\Lambda}=0.6911, \Omega_m=0.3089, \Omega_b=0.0486, H_0=67.74~\mathrm{km}~\mathrm{sec}^{-1}\mathrm{Mpc}^{-1},\sigma_8=0.8159,n_s=0.9667$. Halos are identified  using the friends of friends~(FOF) algorithm~\citep{1985ApJ...292..371D} with a linking length of 0.2 times the mean particle separation. Subhalos are computed at each simulation snapshot using the \texttt{SUBFIND}~\citep{2001MNRAS.328..726S} algorithm.  

The \texttt{BRAHMA} simulations were built to explicitly explore the impact of seeding models on BH populations across cosmic time. To that end, aside from seeding, every other aspect of the galaxy formation model has been kept the same as their predecessor \texttt{Illustris-TNG} simulations~\citep{2018MNRAS.475..676S,2018MNRAS.475..648P,2018MNRAS.475..624N,2018MNRAS.477.1206N,2018MNRAS.480.5113M,2019ComAC...6....2N} which itself is based on the Illustris model~\citep{2013MNRAS.436.3031V, 2014MNRAS.438.1985T}. Here we summarize the key features of the \texttt{Illustris-TNG} galaxy formation model. 

Gas cooling and heating occurs in the presence of a uniform time-dependent UV background. Primordial gas cooling rates~(from $\mathrm{H},\mathrm{H}^{+},\mathrm{He},\mathrm{He}^{+},\mathrm{He}^{++}$) are calculated  based on \citealt{1996ApJS..105...19K}; metal cooling rates are interpolated from pre-calculated tables as in \citealt{2008MNRAS.385.1443S}. Star formation occurs in gas cells with densities exceeding $0.15~\mathrm{cm}^{-3}$ with an associated time scale of $2.2~\mathrm{Gyr}$. These gas cells represent an unresolved multiphase interstellar medium described by an effective equation of state~\citep{2003MNRAS.339..289S,2014MNRAS.444.1518V}. The resulting star particles represent single stellar populations~(SSPs) with an underlying initial mass function adopted from \cite{2003PASP..115..763C}. The SSPs undergo stellar evolution based on \cite{2013MNRAS.436.3031V} with modifications for \texttt{IllustrisTNG} as described in \cite{2018MNRAS.473.4077P}. The resulting chemical enrichment of the SSPs follows the evolution of seven species of metals~(C, N, O, Ne, Mg, Si, Fe) in addition to H and He. These metals pollute the surrounding gas via stellar and Type Ia/II supernova feedback, modelled as galactic scale winds~\citep{2018MNRAS.475..648P}. 

BHs can grow either via accretion or mergers. Gas accretion is modelled by the Eddington-limited Bondi-Hoyle formalism described as 
\begin{eqnarray}
\dot{M}_{\mathrm{bh}}=\mathrm{min}(\dot{M}_{\mathrm{Bondi}}, \dot{M}_{\mathrm{Edd}})\\
\dot{M}_{\mathrm{Bondi}}=\frac{4 \pi G^2 M_{\mathrm{bh}}^2 \rho}{c_s^3}\\
\dot{M}_{\mathrm{Edd}}=\frac{4\pi G M_{\mathrm{bh}} m_p}{\epsilon_r \sigma_T~c}
\label{bondi_eqn}
\end{eqnarray} 
where $G$ is the gravitational constant, $\rho$ is the local gas density, $M_{\mathrm{bh}}$ is the BH mass, $c_s$ is the local sound speed, $m_p$ is the proton mass, and $\sigma_T$ is the Thompson scattering cross section. Accreting BHs radiate at bolometric luminosities given by 
\begin{equation}
    L_{\mathrm{bol}}=\epsilon_r \dot{M}_{\mathrm{bh}} c^2,
    \label{bol_lum_eqn}
\end{equation}
with an assumed radiative efficiency of $\epsilon_r=0.2$. A portion of the radiated energy couples to the surrounding gas as AGN feedback, which is implemented as two distinct modes depending on the Eddington ratio $\eta$. For high Eddington ratios~($\eta > \eta_{\mathrm{crit}}\equiv\mathrm{min}[0.002(M_{\mathrm{BH}}/10^8 \mathrm{M_{\odot}})^2,0.1]$), `thermal feedback' is implemented wherein a fraction of the radiated luminosity is deposited to the gas at a rate of $\epsilon_{f,\mathrm{high}} \epsilon_r \dot{M}_{\mathrm{BH}}c^2$ with $\epsilon_{f,\mathrm{high}} \epsilon_r=0.02$~($\epsilon_{f,\mathrm{high}}$ is the ``high accretion state" coupling efficiency). For low Eddington ratios~($\eta < \eta_{\mathrm{crit}}$), feedback is in the form of kinetic energy that is injected onto the gas at irregular time intervals along a randomly chosen direction. The rate of injection is given by $\epsilon_{f,\mathrm{low}}\dot{M}_{\mathrm{BH}}c^2$ with $\epsilon_{f,\mathrm{low}}$ being the `low accretion state' coupling efficiency~($\epsilon_{f,\mathrm{low}} \lesssim 0.2$). Please see  \cite{2017MNRAS.465.3291W} and \cite{2020MNRAS.493.1888T} for further details. 

We note here that the Bondi prescription is rather simplistic, as it was originally derived for a spherically symmetric accretion flow wherein the gravitational force on the accreting gas is dominated by the BH. In reality, the gas can be self-gravitating~(particularly cold dense gas) and have a substantial angular momentum. Moreover, it has been recently shown that even in the case of spherically symmetric accretion, magnetic fields can suppress the accretion rates by factors of $\sim100$ compared to the Bondi-Hoyle prediction~\citep{2023ApJ...959L..22C}. Nevertheless, our current \texttt{BRAHMA} boxes continue to use Bondi-Hoyle accretion because 1) it has been used in their predecessor \texttt{IllustrisTNG} simulations to successfully reproduce many of the observed galaxy and BH properties, and 2) we want to isolate the impact of seed models on our BH populations. In the near future, we shall be exploring our seed models with alternate accretion models~(Weinberger et al in prep, Burger et al in prep). 

Finally, the BH dynamics cannot be naturally resolved in our simulations since the DM particles are $\sim10$ times larger than the BHs themselves. To prevent artificially large kicks from DM particles, we ``re-position" the BHs to the nearest potential minimum within their ``neighborhood"~(defined by $n^{\mathrm{gas}}_{\mathrm{ngb}}=64$ nearest neighboring gas cells). BHs are promptly merged when at least one of them is within the ``neighbor search radius"~($R_{\mathrm{Hsml}}$ comprising of $n^{\mathrm{gas}}_{\mathrm{ngb}}$ gas cells) of the other.

\subsection{BH seeding models and the resulting simulation suite}
The key novel feature of \texttt{BRAHMA} is the implementation of new physically motivated prescriptions for BH seeding. These prescriptions were originally developed and tested in our previous work using high resolution zoom simulations \citep{2021MNRAS.507.2012B,2022MNRAS.510..177B,2024MNRAS.529.3768B}, and subsequently implemented in the uniform suite of \texttt{BRAHMA} simulations~\citep{2024MNRAS.531.4311B,2024MNRAS.533.1907B}. The existing \texttt{BRAHMA} suite consists of $>15$ uniform boxes wherein we systematically explore a large ensemble of seed models that span a wide range of seeding environments as well as seed masses ranging from $\sim10^3-10^6~M_{\odot}$. 

In this paper, we select a subset of $[18~\mathrm{Mpc}]^3$ \texttt{BRAHMA} boxes that were run to $z=0$ using $512^3$ dark matter~(DM) particles and equal number of initial gas cells. The resulting DM mass resolution $1.5\times10^6~M_{\odot}$ and the gas mass and stellar mass resolution is $\sim10^5~M_{\odot}$. The comoving gravitational softening length is $0.72~\rm kpc$. These boxes seed BHs close to their gas mass resolution; i.e. $1.5\times10^5~M_{\odot}$. Depending on our seed models, these initial BHs can be interpreted either as being directly formed as \textit{heavy} seeds, or as \textit{extrapolated descendants} of unresolved lower mass seeds~(as explained below).

\subsubsection{Heavy seed models:}

To model heavy seeds, we use distinct combinations of the following four seeding criteria that are motivated by presumed feasible environments for DCBH formation as first proposed by \cite{2006MNRAS.371.1813L}:

\begin{itemize}
    \item \textit{Dense $\&$ metal-poor gas mass criterion:} When this criterion is applied, seeds are able to form only in halos that exceed a critical gas mass threshold that is simultaneously dense enough to form stars~($\gtrsim0.15~\rm cm^{-3}$), and yet devoid of metals~($Z\lesssim10^{-4}~Z_{\odot}$). There is no  obvious choice for what is the best value for this threshold, and we explored a range of possibilities from $\sim5-150~\seedmass$ in \cite{2021MNRAS.507.2012B} using zoom simulations run till $z=7$. The variations in the overall seed abundances and $z=7$ BH masses were by factors of $\sim10$. In this work, we choose a lenient threshold of $5~\seedmass$, such that we could thereafter stack additional seeding criteria~(described below) and still form sufficient numbers of seeds within our simulation volume. 
    \item \textit{LW flux criterion:} This criterion further requires the dense $\&$ metal-poor gas mass to be exposed to a minimum Lyman Werner~(LW) flux of $10~J_{21}$. In the absence of a full radiative transfer calculation, the local LW intensity was computed using an analytical formalism adopted from \cite{2014MNRAS.442.2036D}.

    \textit{Halo mass criterion:} We only allow seeding in sufficiently resolved halos~(with $>32$ DM particles). Therefore, the resolution limit of our simulations also implicitly imposes an additional halo mass threshold of $4.8\times10^7~M_{\odot}$ for seeding. Note that this threshold is well within the atomic cooling limit that evolves from $\sim10^7~M_{\odot}$ to $10^8~M_{\odot}$ from $z\sim20$ to $0$. One could use higher resolution simulations to explore the formation of heavy seeds in even smaller halos, but that would require additional physics that are not included in our simulations~(for e.g. $H_2$ cooling). We shall explore this in future work.
    
    \item \textit{Gas-spin criterion:} This criterion restricts seeding to halos with a gas spin that is below the Toomre instability threshold. The adoption of this criterion was based on \cite{2006MNRAS.371.1813L} and the detailed description is provided in \cite{2022MNRAS.510..177B}.
    \item \textit{Halo environment criterion:} This criterion allows seeding to only occur in halos in rich environments; i.e. they have at least one neighbor of comparable or higher mass within a distance of 5 times the virial radius. This is intended the capture the impact of dynamical heating due to mergers of massive halos, which has been shown to substantially alleviate the stringent requirement of strong LW fluxes for DCBH formation~\citep{2019Natur.566...85W,2020OJAp....3E..15R}.
\end{itemize}

We use the same four heavy seed model simulation boxes as \cite{2024MNRAS.533.1907B}. The seed models in these boxes were constructed by incrementally stacking the above seeding criteria: The first box~(\texttt{SM5}) only applies the \textit{dense \& metal-poor gas mass criterion} and the implicitly applied \textit{halo mass criterion}. The second box~(\texttt{SM5_LW10}) additionally applies the \textit{LW flux criterion}. The third box~(\texttt{SM5_LW10_LOWSPIN}) further adds the \textit{gas-spin criterion}. The fourth box~(\texttt{SM5_LW10_LOWSPIN_RICH}) also includes the \textit{halo environment criterion}. Since these models assume that the initial $1.5\times10^5~M_{\odot}$ BHs were directly formed as heavy seeds, we shall hereafter refer to them as \textit{direct heavy seeds} or ``DHS". We also note that these DHS models will produce much more numerous heavy seeds compared to canonical DCBH formation scenarios discussed in the literature that require extremely high critical LW fluxes~($\gtrsim1000~J_{21}$). Based on the results from our previous work~\citep{2022MNRAS.510..177B}, we expect an extremely small number of seeds~(if any) to be produced in our boxes at such high critical LW fluxes. Therefore, we don't include these very restrictive DCBH formation scenarios in this paper, as we don't expect them to be the origins for the bulk of the locally observed SMBH populations.

\subsubsection{Low-mass seed models:}
Low-mass seeds are generally challenging to model in cosmological simulations due to the difficulty in directly resolving them. To that end, in \cite{2024MNRAS.529.3768B}, we developed a novel stochastic seed model that can faithfully capture the higher mass descendants of seeds that are $\sim10-100$ times below the simulation mass resolution. Since the initial BHs in this case are descendants of the \textit{actual} seeds, we refer to them as \textit{extrapolated seed descendants} or ESDs. These models are directly calibrated using highest resolution simulations that explicitly resolve our target low-mass seeds. In this work, our target seed mass is $2.2\times10^3~M_{\odot}$. To ensure that our $1.5\times10^5~M_{\odot}$ ESDs faithfully represent descendants of $\sim10^3~M_{\odot}$ seeds, the stochastic seed model uses the following three seeding criteria:

\begin{itemize}

\item \textit{Stochastic galaxy mass criterion}: When this criterion is applied, ESDs are formed in galaxies\footnote{Note here that \textit{galaxies} here are essentially identified as friends-of-friends groups, but they have $1/3$ of the standard linking-length compared to halos. We refer to these objects as ``best friends-of-friends" or bFOFs. We place ESDs in bFOFs instead of FOFs to accomodate the possibility of multiple seed descendants per halo.} over a broad distribution of total masses~(star+gas+DM) that evolves with redshift. This criterion is required because in \citealt{2024MNRAS.529.3768B,2024MNRAS.531.4311B}, descendants of $2.2\times10^3~M_{\odot}$ seeds were found to assemble in galaxies spanning a broad distribution of masses.

\item \textit{Stochastic galaxy environment criterion}: This criterion preferentially forms ESDs in galaxies living in rich environments with a higher number of comparable or more massive neighbors. The need for this criterion arose because of the merger-dominated growth of $2.2\times10^3~M_{\odot}$ seeds at $z\gtrsim3$~(as we found in \citealt{2024MNRAS.529.3768B,2024MNRAS.531.4311B}). More specifically, the mergers lead to preferentially higher BH growth in galaxies living in rich environments~(at fixed galaxy mass) since they had a more extensive merger history.

\item \textit{Unresolved minor mergers}: Since our simulations cannot capture BH growth due to mergers where the primary BH is resolvable but the secondary BH is not~(i.e. the secondary BH mass is lower than the ESD mass), their contribution is explicitly added to the simulations\footnote{The simulations that use the stochastic seed model also do not explicitly resolve mergers wherein \textit{both} primary and secondary BHs are below the ESD mass. However, their contributions are implicitly included in the creation of ESDs via the \textit{stochastic galaxy mass and environment criterion}.}. More specifically, for every resolved merger event, we add an additional contribution equal to 4 times the ESD mass to account for mass growth due to unresolved minor mergers. This contribution was calibrated based on results from high resolution simulations of \cite{2024MNRAS.531.4311B} that explicitly resolve these minor mergers.  
\end{itemize}

Key inputs for our stochastic seed model include the shape and the time-evolution of the galaxy mass distribution, seed probability as a function of galaxy neighbor counts, and the BH growth rate due to unresolved minor mergers. These inputs are provided by high resolution simulations\footnote{These are 9 Mpc box simulations that resolve the gas down to $\sim10^3~M_{\odot}$ resolution. They are referred to as \texttt{BRAHMA-9-D3} simulations in \citealt{2024MNRAS.531.4311B}.} that explicitly resolve our target seed mass of $2.2\times10^3~M_{\odot}$. In other words, the stochastic seed model is designed to ensure that ESDs are placed in those environments wherein the high resolution simulations would naturally assemble descendant BHs comparable to the ESD mass. For full details of the modeling and calibration of the stochastic seed model, please refer to \cite{2024MNRAS.529.3768B}. Our fifth and final simulation box used in this work applied this stochastic seed model and is labeled as ``\texttt{ESD:STOCHASTIC}".   

To summarize, our simulation suite comprises of five $[18~\mathrm{Mpc}]^3$ boxes that initialize BHs at $1.5\times10^5~M_{\odot}$. The first four boxes apply the DHS seed models, namely \texttt{SM5}, \texttt{SM5_LW10}, \texttt{SM5_LW10_LOWSPIN}, and \texttt{SM5_LW10_LOWSPIN_RICH}, wherein the initial BHs can be interpreted as heavy seeds born out of direct collapse. In the fifth \texttt{ESD:STOCHASTIC}
 box, the initial BHs are interpreted as descendants of lower mass $\sim10^3~M_{\odot}$ seeds. Note that for these relatively small boxes, the results may be impacted by missing large scale modes as well as cosmic variance. However, as we shall see later, our predicted local BH mass functions are similar to those from the much larger \texttt{IllustrisTNG} simulations for $\gtrsim10^7~M_{\odot}$ BHs. This indicates that at least for the quantities we are interested in, such as mass functions and occupation fractions of $\sim10^5-10^8~M_{\odot}$ BHs, the results are not substantially impacted by our limited volume. Lastly, these small volumes are not able to probe the high mass end of the galaxy and BH populations. However, our key focus here is on  the low mass end, particularly the local dwarf galaxies, which these simulations can effectively probe.    

\section{Results}
\label{Results sec}

\begin{figure*}

\includegraphics[width=8 cm]{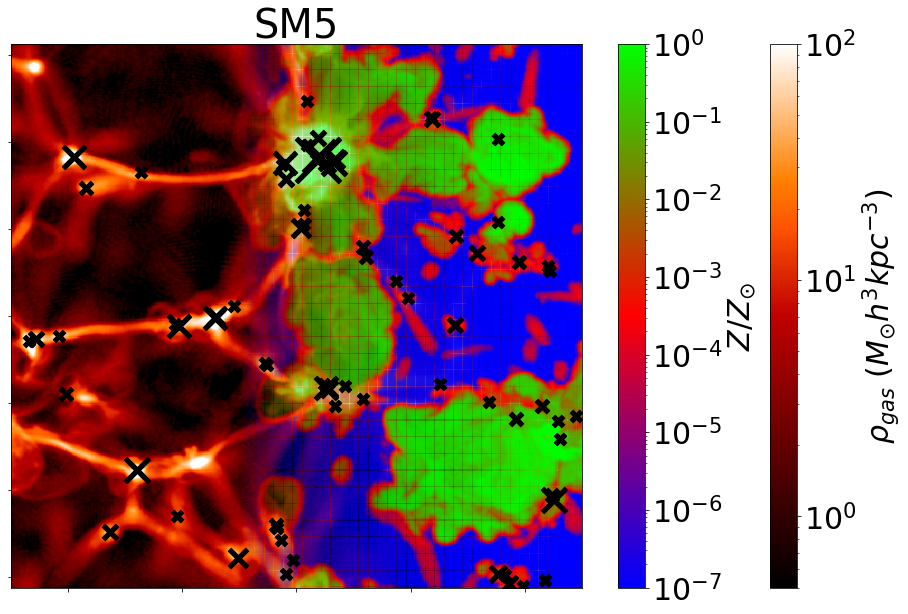}
\includegraphics[width=8 cm]{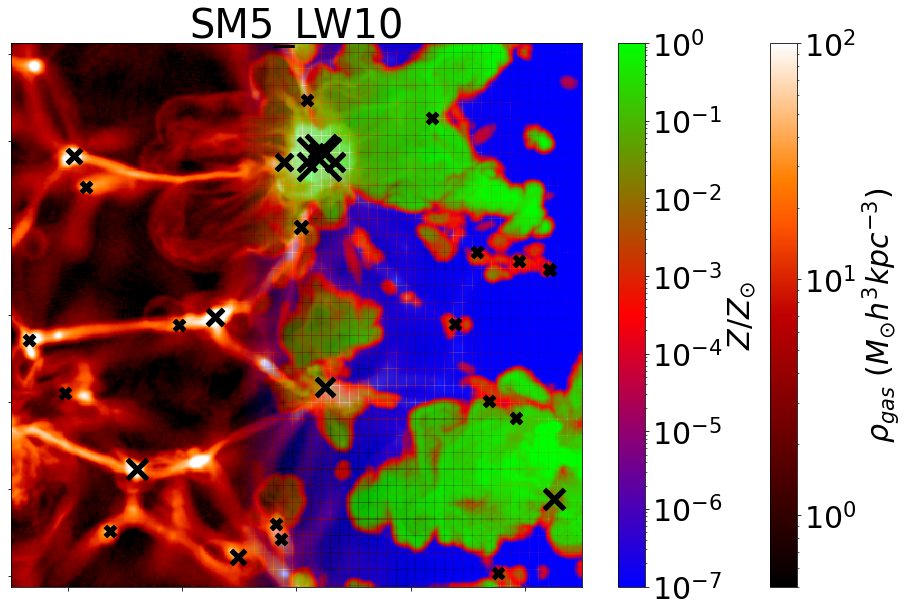} \\
\includegraphics[width=8 cm]{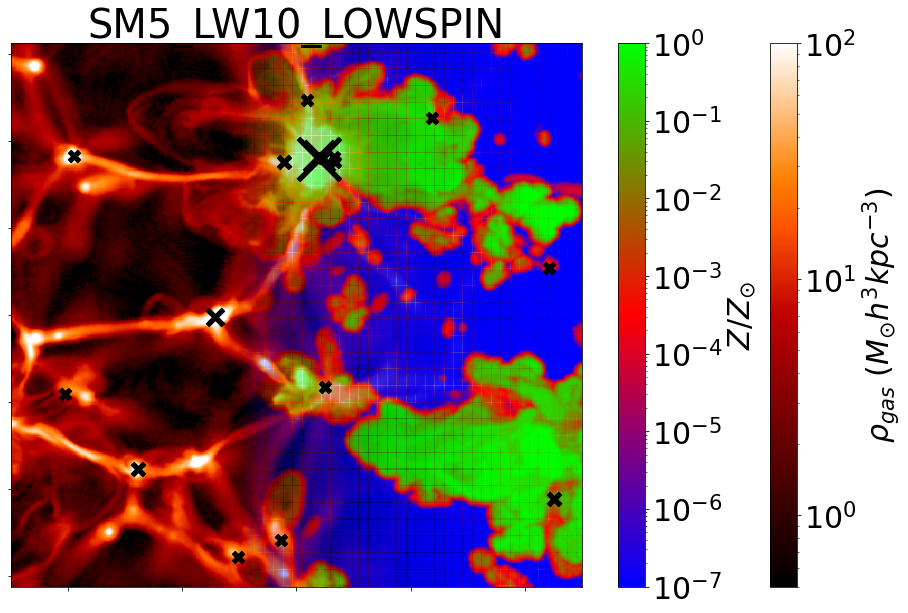}
\includegraphics[width=8 cm]{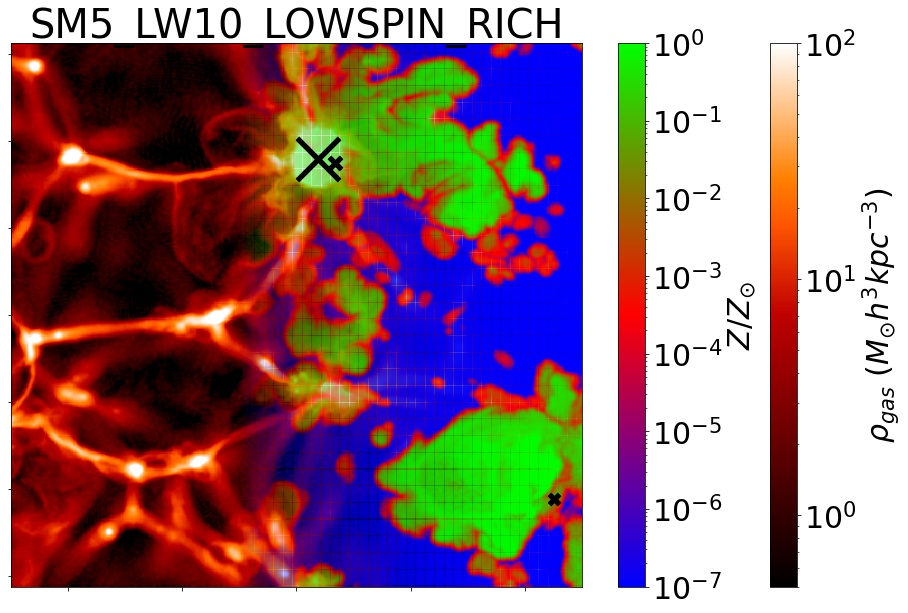}

\includegraphics[width=8 cm]{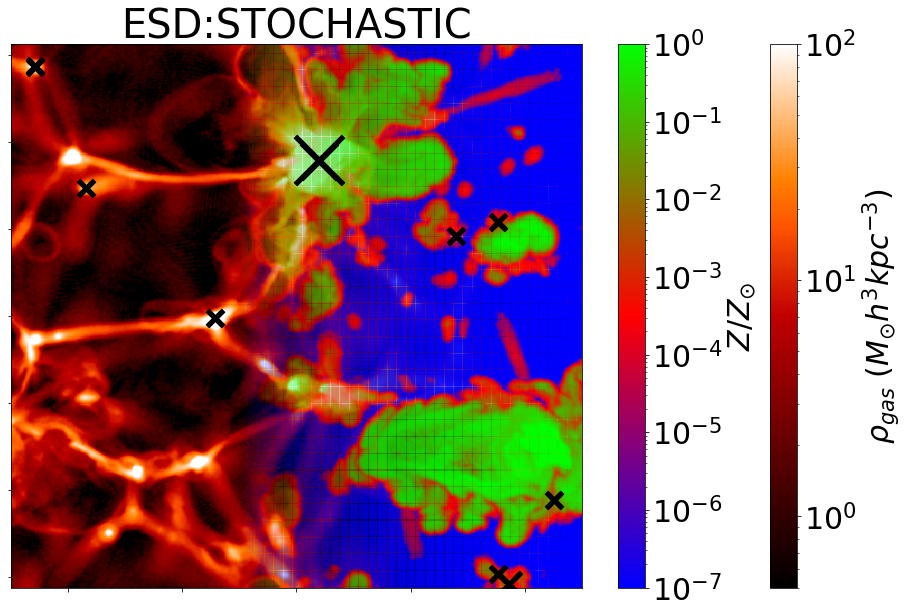}

\caption{Visualizations of the $z=0$ universe in our $[18~\mathrm{Mpc}]^3$ simulation boxes that use the different seed models with an initial mass of $1.5\times10^5~M_{\odot}$. The left side of the boxes shows the gas density profiles and the right side shows the gas metallicity profiles at $z=0$, averaged over a $200~\mathrm{kpc}$ thick slice along the line of sight coordinate. The top four boxes~(\texttt{SM5}, \texttt{SM5_LW10}, \texttt{SM5_LW10_LOWSPIN} and  \texttt{SM5_LW10_LOWSPIN_RICH}) show four models that assume the initial BHs are directly formed as heavy seeds~(hereafter ''direct heavy seeds" or ``DHS"). The lowermost box~(\texttt{ESD:STOCHASTIC}) uses a stochastic seed model built in \protect\cite{2024MNRAS.529.3768B} that assumes that the initial BHs are descendants of lower mass seeds~(hereafter ''extrapolated seed descendants" or ``ESDs").  BHs are represented by the black crosses with sizes commensurate to their masses. All the seed models produce the most massive BHs in the most overdense regions at the intersections of filaments. However, the more lenient seed models produce much larger numbers of lower mass BHs compared to the stricter seed models.}
\label{visualization}
\end{figure*}

We examine a range of properties that describe the $z=0$ BH populations with the goal of identifying local signatures of seeding models. Figure \ref{visualization} shows a visualization of the spatial distributions of $z=0$ BH populations throughout the simulation volumes, plotted on top of the 2D gas density and metallicity profiles. For the most lenient DHS model i.e. \texttt{SM5}, the most massive BHs are formed in the most overdense regions that are typically at the intersections of filaments. In addition, there is a more abundant population of lower mass BHs in less overdense regions, typically along the filaments. As we make the DHS models more restrictive, the lower mass BHs residing along the filaments are suppressed whereas the most massive BHs remain virtually unaffected. The ESD model contains substantially fewer numbers of BHs than the most lenient DHS model, but a larger number of BHs than the most restrictive DHS model. However, note that here we are only counting the number of $\gtrsim10^5~M_{\odot}$ BHs whereas the ESD model is meant to represent a Universe that should also contain a much larger number of $\sim10^3~M_{\odot}$ seeds that cannot be resolved in our boxes. Overall, the above results hint at the survival of robust signatures of seeding at the low mass end of the BH population, which we quantify in more detail in the following subsections. 

\subsection{Seed formation history}

\begin{figure}
\includegraphics[width=8 cm]{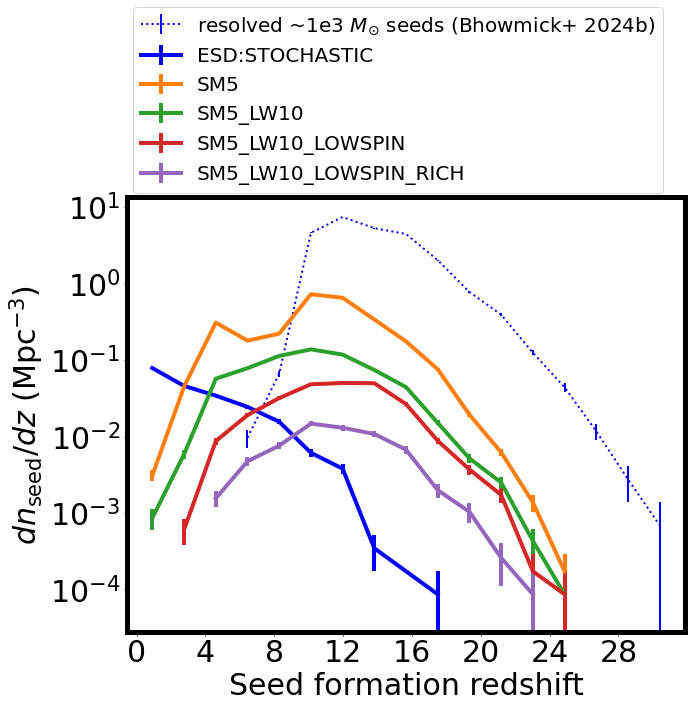}

\caption{The comoving number density of new seeds that form at various redshifts. Predictions for the four DHS models: \texttt{SM5}, \texttt{SM5_LW10}, \texttt{SM5_LW10_LOWSPIN} and  \texttt{SM5_LW10_LOWSPIN_RICH}, are shown as solid lines in orange, green, red and purple respectively. The blue solid line shows the ESD model~(\texttt{ESD:STOCHASTIC}) that seeds $1.5\times10^{5}~M_{\odot}$ BHs as if they are descendants of unresolved lower mass $\sim10^{3}~M_{\odot}$ seeds. The ESD model was calibrated from higher resolution simulations that explicitly resolve the $\sim10^{3}~M_{\odot}$ seeds; the dotted line shows the number density and halo mass distributions of the actual $\sim10^{3}~M_{\odot}$ seeds that were fully resolved in a $[9~\mathrm{Mpc}]^3$ high resolution box from \protect\cite{2024MNRAS.531.4311B}.  Therefore, the dotted and solid blue lines are meant to represent the same underlying physical seed model. The foregoing color scheme for the different boxes is the same for all the figures hereafter. While DHSs emerge earlier than the ESDs, their formation is suppressed by metal-enrichment at $z\lesssim10$. On the other hand, even though ESDs emerge later, they continue to form in higher numbers at decreasing redshifts, all the way to $z\sim0$.}
\label{seed_formation}
\end{figure}

\begin{figure}
\includegraphics[width=8 cm]{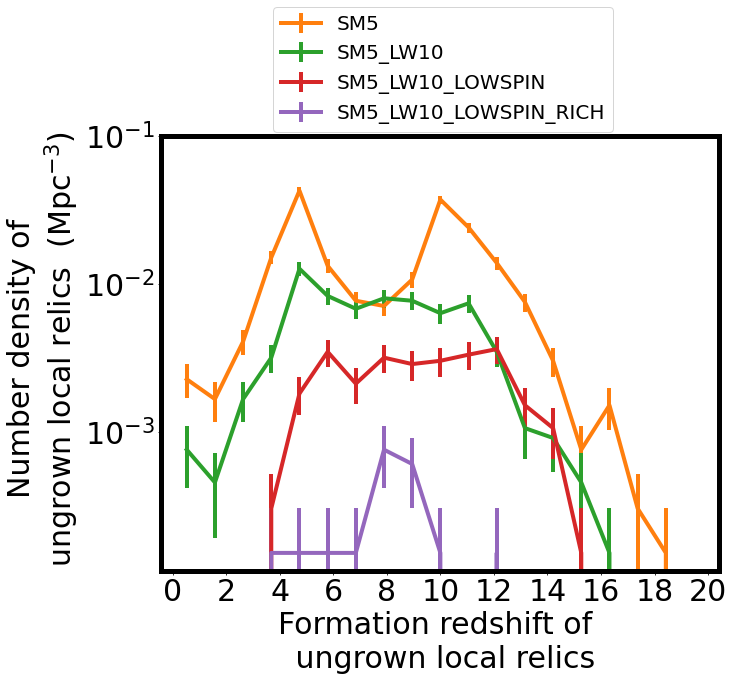}

\caption{For our four DHS models, we show the distribution~(number density per unit redshift bin) of the formation redshifts of seeds that did not undergo any significant growth via accretion or mergers all the way down to $z=0$. Specifically, we select seeds that only grew by $\lesssim10~\%$ of their initial mass. We find that in all four DHS models, most of these ungrown seeds were formed in the Universe between $z\sim5-10$.}  
\label{local_relics}
\end{figure}

While this paper focuses on $z=0$, it is imperative to first look at the overall seed formation history in order to contextualize the results in the local Universe. Figure \ref{seed_formation} shows the rate at which new seeds are forming per unit redshift. For the four DHS models, the first episodes of seed formation occur around $z\sim24$ as the first regions of dense~(star forming) gas start to emerge. As the redshift decreases, seed formation continues to increase until it peaks at $z\sim10$. After $z\sim10$, this mode of seed formation slows down due to metal enrichment. However, even halos that are undergoing metal enrichment, can~(and do) form seeds. In fact, as demonstrated in our previous papers~\citep{2022MNRAS.510..177B,2024MNRAS.529.3768B,2024MNRAS.531.4311B} most of our seeds form in transient pockets of dense and metal poor gas in partially metal enriched halos. Therefore, in the two most lenient DHS models~(\texttt{SM5} and \texttt{SM5_LW10}), there is some amount of seed formation that continues to occur all the way to $z=0$. For the two more restrictive DHS models, \texttt{SM5_LW10_LOWSPIN} and \texttt{SM5_LW10_LOWSPIN_RICH}, the last seeds form around $z\sim3$ and $z\sim5$ respectively; however, even for these models, some amount of seed formation could have happened at lower redshifts if our boxes were larger. 

The seeds that form very close to $z\sim0$ do not have enough time to grow into higher mass BHs. Therefore, these seeds will populate the lowest mass end of the local BH mass functions. However, the low mass end of the local BH mass functions will also be populated by seeds that form at higher redshifts but do not grow appreciably via either accretion or mergers by $z\sim0$. In Figure \ref{local_relics}, we take the four DHS models and plot the distribution of formation redshifts for those seeds that undergo negligible growth~($\lesssim10~\%$ of their initial seed mass) by $z\sim0$. We can readily see that for all DHS models, the majority of the ungrown BHs are local relics of seeds that formed between $z\sim5-10$. While these ungrown seeds do~(unsurprisingly) form at lower redshifts compared to when most seeds form in general~($z\sim12$), they still form at relatively high redshift compared to $z=0$. We show in Sections \ref{BH number density evolution sec} and \ref{BH mass functions sec} that these local relics of higher redshift seeds~($z\sim5-10$) essentially dominate the lowest mass end of our simulated local BH populations, and therefore retain strong signatures of BH seeding in their overall abundances.

Interestingly, for the $\texttt{SM5}$ model, the distributions in Figure \ref{local_relics}~(orange lines) have two peaks at $z\sim10$ and $z\sim5$. While the peak at $z\sim10$ coincides with the peak in the overall seed formation history shown in Figure \ref{seed_formation}, the second peak at $z\sim5$ is caused by a noticeable temporary phase of enhancement in the overall seed formation rates around $z\sim5$~(orange lines in Figure \ref{seed_formation}). This enhancement likely occurs due to an emergence of a fresh new set of seed forming halos~(with enough dense and metal-poor gas mass) in less extreme peaks of the initial overdensity field, that somehow avoided being completely polluted by metals from prior episodes of metal enrichment throughout the box. We see hints of similar enhancements in seed formation at $z\sim5$ even in the more restrictive DHS models, but they are much less pronounced compared to the \texttt{SM5} model.


For the ESD model~(blue line in Figure \ref{seed_formation}), the (extrapolated) $1.5\times10^5~M_{\odot}$ seeds start to form at $z\sim16$, which is significantly delayed compared to the DHS models. Recall again that the ESDs are meant to be descendants of unresolved lower mass $\sim10^3~M_{\odot}$ seeds. The blue dotted line in Figure \ref{seed_formation} shows the seed formation history of the $\sim10^3~M_{\odot}$ seeds from one of the high resolution boxes of \cite{2024MNRAS.531.4311B} where they were explicitly resolved~(the ESD model was calibrated using this simulation). In this simulation, the $2.2\times10^3~M_{\odot}$ seeds were placed in halos with sufficient dense and metal-poor gas mass ($>10^4~M_{\odot}$) and total mass~($>6\times10^6~M_{\odot}$). As we can see, the $\sim10^3~M_{\odot}$ seeds form in much higher numbers~(per unit volume) than any of the DHS models. Despite this, their higher mass $\sim10^5~M_{\odot}$ descendants start forming significantly later than the DHS models. In other words, it takes longer to grow  $\sim10^5~M_{\odot}$ BHs from lower mass seeds, compared to directly producing the $\sim10^5~M_{\odot}$ BHs as heavy seeds using our DHS models. However, while seed-formation in our DHS models is slowed down by metal-enrichment at $z\gtrsim12$, the ESD model continues to  produce an increasing number of $\sim10^5~M_{\odot}$ ESDs down to much lower redshifts. 

\subsection{BH number density evolution}
\label{BH number density evolution sec}

\begin{figure*}
\includegraphics[width=13 cm]{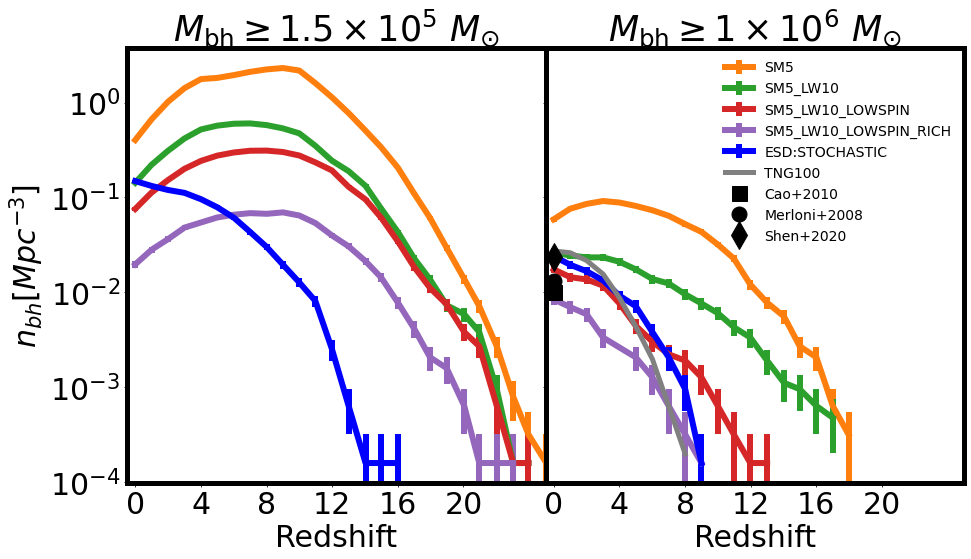}
\caption{Comoving number density evolution of $\gtrsim10^5$ and $\gtrsim10^6~M_{\odot}$ BHs are shown in the left and right panels respectively. Black data points show the local observational constraints from \protect\cite{2008MNRAS.388.1011M}, \protect\cite{2010ApJ...725..388C} and \protect\cite{2020MNRAS.495.3252S}. The grey line shows the prediction from the TNG100 simulation. The $z=0$ number densities exhibit significant seed model variations for both $\gtrsim10^5$ and $\gtrsim10^6~M_{\odot}$ BHs. }
\label{Number_density}
\end{figure*}

\begin{figure*}
\includegraphics[width=16 cm]{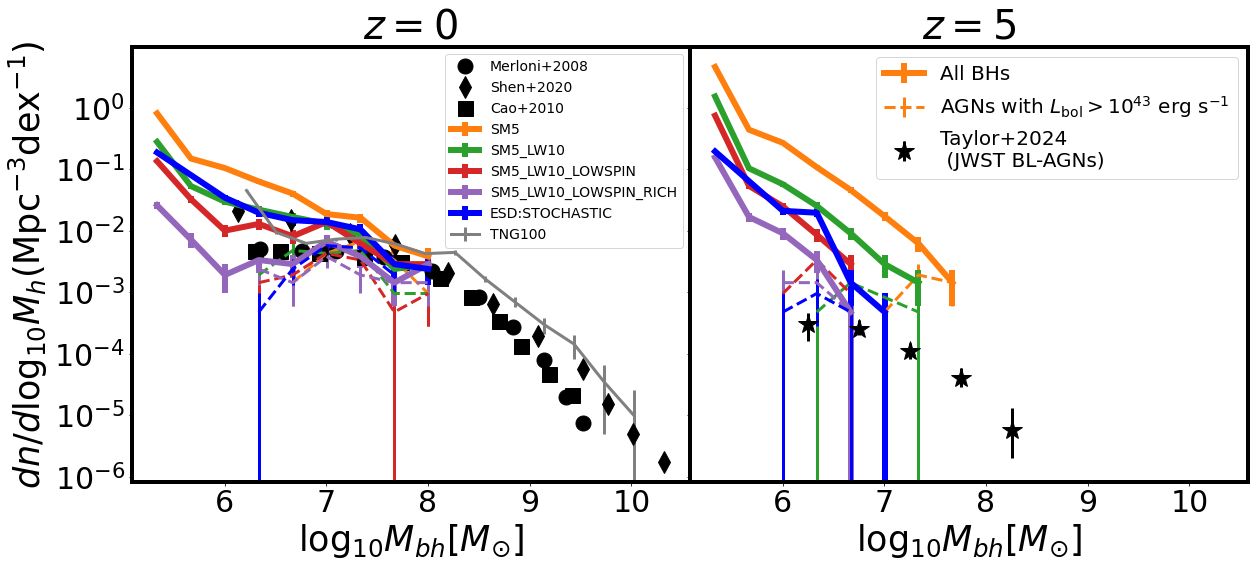}
\caption{BH mass functions~(solid lines) at $z=0$~(left) and $z=5$~(right) for the different seed models as compared to local observations~(black data points) from \protect\cite{2008MNRAS.388.1011M}, \protect\cite{2010ApJ...725..388C}, \protect\cite{2020MNRAS.495.3252S}. At $z=5$, we show the recent constraints from JWST BL-AGNs from \protect\cite{2024arXiv240906772T}; however, note that they only include active AGNs above the JWST detection limit. The dashed lines show the contributions from active BHs with $L_{\rm bol}>10^{43}~\mathrm{erg~s^{-1}}$. The thin grey line shows the prediction from TNG100 simulation. At $z=0$, the signatures of seeding are strongest for $\sim10^5-10^{6}~M_{\odot}$ BHs. They persist at masses of $\sim10^6-10^{7}~M_{\odot}$, but eventually disappear at $\gtrsim10^{7}~M_{\odot}$. In contrast, at $z=5$, the signatures of seeding exists at all BH masses~($\sim10^5-10^8~M_{\odot}$) probed by our boxes.}
\label{mass_function}
\end{figure*}

Figure \ref{Number_density} shows the number density evolution of $\gtrsim10^5~M_{\odot}$ and $\gtrsim10^6~M_{\odot}$ BHs. Let us first focus on $\gtrsim10^5~M_{\odot}$ BHs~(left panel). As expected from the seed formation history, in the DHS models, the number density of $\gtrsim10^5~M_{\odot}$ BHs steadily increases with time between $z\sim24-10$. At $z\lesssim10$ when new seed formation starts to drop, the number density steadily decreases from $z\sim10-0$. This is because the existing seeds start to merge with one another. At $z=0$, the number densities predicted by the DHS models vary from $0.02-0.4~\mathrm{Mpc}^{-3}$. Recall from the previous section that these number densities are essentially dominated by the local ungrown relics of high-redshift~($z\sim5-10$) seeds. The ESD seed model, because of its distinct seed formation history, also has a distinct number density evolution compared to the DHS models. While the number density decreases from $z\sim12$ to $z\sim0$ in the DHS models, it steadily increases from $z\sim16$ to $z\sim0$ in the ESD model. Consequently, despite having substantially smaller number densities compared to the DHS models at $z\gtrsim7$, the ESD model catches up with the DHS models at $z=0$. The ESD model predicts a $z=0$ number density of $\sim0.1~\mathrm{Mpc}^{-3}$ for $\gtrsim10^5~M_{\odot}$ BHs, which is similar to the two intermediary DHS models~(\texttt{SM5_LW10} and \texttt{SM5_LW10_LOWSPIN}). This is $\sim8$ times larger than the most restrictive DHS model~(\texttt{SM5_LW10_LOWSPIN_RICH}), while being $\sim2$ times fewer than the most lenient DHS model~(\texttt{SM5}). Overall, between the ESD model and the four DHS models, variations in the number densities of $\gtrsim10^5~M_{\odot}$ BHs can be up to factors of $\sim20$. 
This implies that future measurements of the number density of $\gtrsim10^5~M_{\odot}$ BHs in the local Universe, could provide strong constraints for seed models. Now we focus on the number density of $\gtrsim10^6~M_{\odot}$ BHs~(right panel of Figure \ref{Number_density}). At $z=5$, the seed model variations in the number density span more than two orders of magnitude from a few times $\sim10^{-4}$ to $\sim6\times10^{-2}~\mathrm{Mpc}^{-3}$. As we get to lower redshifts, the seed model variations start to become smaller. By $z=0$, the number density of $\gtrsim10^6~M_{\odot}$ BHs varies by a factor of $\sim5$, ranging between $\sim0.01-0.05~\mathrm{Mpc}^{-3}$. Therefore, with precise enough observational measurements, even the abundances of $\gtrsim10^6~M_{\odot}$ BHs could potentially be used for constraining seed models. However, as we discuss in Section \ref{caveats}, possible degeneracies with signatures from other aspects of BH physics, such as accretion, dynamics and feedback, may pose a significant challenge.

The current samples of local $\gtrsim10^6~M_{\odot}$ SMBH populations are unlikely to be complete. Nevertheless, it has been possible to estimate the local BH mass functions~(discussed in the next section) using the observed scaling relations between BH mass vs. galaxy luminosities and stellar velocity dispersion, combined with the local galaxy luminosity and velocity functions~\citep{2004MNRAS.351..169M}. The different estimates of the BH mass functions~\citep{2008MNRAS.388.1011M,2009ApJ...690...20S,2010ApJ...725..388C,2020MNRAS.495.3252S} in the literature have some spread that depends on the exact scaling relations they use~(see \citealt{2009ApJ...690...20S} for details). We integrate these BH mass functions to obtain the number density of $\gtrsim10^6~M_{\odot}$ BHs and show them as black data points in Figure \ref{Number_density}. Notably, the ESD model as well as the two intermediary DHS models produce local number densities that agree well with the latest observational constraints by \cite{2020MNRAS.495.3252S}. The most restrictive DHS model~(\texttt{SM5_LW10_LOWSPIN_RICH}) prediction is slightly lower than the estimate from \cite{2020MNRAS.495.3252S} by a factor of $\sim3$, but is close to the estimates of  \cite{2008MNRAS.388.1011M} and \cite{2010ApJ...725..388C}. At the other end, the prediction of the most lenient DHS model~(\texttt{SM5}) lies a factor of $\sim3$ above the \cite{2020MNRAS.495.3252S} measurements. However, as we mentioned earlier, the uncertainties in these estimates may be significant as the scaling relations from which they are derived do not arise from a ``complete" population of local $\gtrsim 10^6~M_{\odot}$ SMBHs. With more precise estimates of the BH number densities in the future, we will be able to make a better assessment about which seed models are more viable than others.  



\subsection{BH mass functions}
\label{BH mass functions sec}

The left panel of Figure \ref{mass_function} shows the BH mass functions (BHMFs) at $z=0$ compared against the observational estimates for which the number densities were computed in the previous figure~\citep{2008MNRAS.388.1011M,2010ApJ...725..388C,2020MNRAS.495.3252S}. As we already saw in Figure \ref{Number_density}, the seed model variations are the strongest for the lowest mass $\sim10^5-10^6~M_{\odot}$ BHs. Recall again that for the DHS models, the smallest $\sim10^5~M_{\odot}$ BHs are predominantly composed of local ungrown relics of seeds that formed at high-redshift~($z\sim5-10$). The models start to gradually converge as we go from $\sim10^6-10^7~M_{\odot}$, until they become similar to one another for $\gtrsim10^7~M_{\odot}$ BHs. In fact, even the TNG100 simulation, which has a very different seed model~($\sim10^6~M_{\odot}$ seeds are placed in $\sim10^{10}~M_{\odot}$ halos), predicts a very similar BHMF to the \texttt{BRAHMA} boxes at $\gtrsim10^7~M_{\odot}$. All the seed models are also consistent with the observational BHMFs at $M_{bh}\gtrsim10^7~M_{\odot}$. 

For $M_{bh}\sim10^6-10^7~M_{\odot}$ BHs, the comparison between our simulated and the observed BHMFs is naturally similar to what we saw in the BH number densities of $\gtrsim10^6~M_{\odot}$ in Figure \ref{Number_density}. The most lenient DHS model is slightly higher than the latest \cite{2020MNRAS.495.3252S} BHMF and the most restrictive DHS model prediction is close to the measurements from  \cite{2008MNRAS.388.1011M} and \cite{2010ApJ...725..388C} BHMF. The remaining models~(\texttt{ESD:STOCHASTIC}, \texttt{SM5_LW10} and \texttt{SM5_LW10_LOWSPIN}) are closest to the \cite{2020MNRAS.495.3252S} BHMF. Here again, we recall that in the absence of a complete population of local SMBHs down to $\sim10^6~M_{\odot}$, these observational BHMFs may be subject to significant modeling uncertainties from the underlying BH-galaxy scaling relations they were derived from. We expect these measurements to converge in the future and potentially rule out some of our seeding scenarios~(modulo uncertainties in BH accretion, dynamics and feedback modeling).

Notably, the predicted seed model variations in the $z=0$ BHMFs are qualitatively distinct compared to high-z BHMFs. While the $z=0$ BHMFs for the different seed models converge at the massive end, the $z=5$ BHMFs~(right panel of Figure \ref{mass_function}) tend to have stronger seed model variations at the massive end. We also reported this in \cite{2024MNRAS.531.4311B} using lower mass seed models~($\sim10^3~M_{\odot}$). The reason for the difference in behavior at low-z vs. high-z lies within the relative importances of the two distinct modes of BH growth; i.e. BH-BH mergers vs gas accretion. 

\begin{figure*}
\includegraphics[width=18 cm]{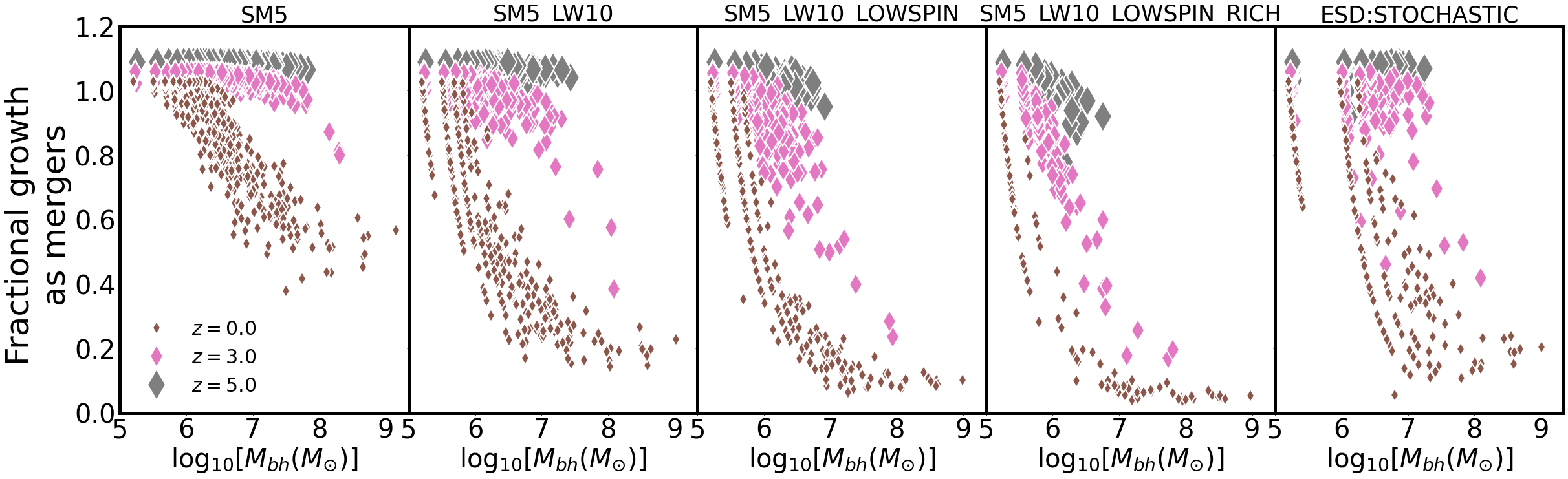}

\caption{The fraction of the accumulated BH mass that is contributed via mergers~(rather than gas accretion) is plotted as a function of BH mass. The different colors show BH populations at $z=0,3~\&~5$ snapshots. We added artificial y-axis offsets for $z=3~\&~5$ for clarity. At $z=5$~(and higher redshifts), BHs grow predominantly via mergers. The importance of accretion increases as we go to lower redshifts. Even at $z=0$, the BH growth between $\sim10^5$ to $10^6~M_{\odot}$ is merger dominated. For more massive local BHs~($M_{bh}\gtrsim10^7~M_{\odot}$) however, accretion is the dominant contributor to the BH mass assembly. The signatures of seeding tend to most strongly persist in regimes wherein the mergers dominate BH growth. Accretion tends to erase signatures of seeding.}
\label{mergers_vs_accretion}
\end{figure*}

In Figure \ref{mergers_vs_accretion}, we show the fraction of BH mass growth that occurs due to mergers~(as a function of BH mass) for BH populations at redshifts $5,3~\&~0$. At $z=5$ (and higher redshifts), the BH growth is largely dominated by mergers, as also discussed in \citealt{2024MNRAS.529.3768B,2024MNRAS.531.4311B,2024MNRAS.533.1907B}. This is primarily because stellar feedback prevents the gas from accumulating at the halo centers to fuel BH accretion. The $M_{bh}^2$ scaling of the Bondi accretion rate also contributes to reduced accretion rates for seed BHs. At lower redshifts, the contribution from accretion becomes increasingly important particularly for higher mass BHs. At $z=0$, accretion is the dominant contributor to BH growth for $\gtrsim10^7~M_{\odot}$ BHs for all the seed models except the most lenient DHS model~(where the contribution from mergers and accretion are roughly equal). All that being said, when BHs grow via mergers, seed model variations tend to be stronger for more massive BHs~(as in $z=5$); this is because the merger-driven BH growth relies on the availability of seeds to fuel the mergers~(that strongly depends on the underlying seed model). On the other hand, when BHs grow via gas accretion~(as in $z=0$), seed model variations tend to be weaker because accretion-driven BH growth relies on the availability of gas around the BH. In this case, only a select few BHs living in certain gas rich environments will grow, regardless of the seed models.

Note also that even at $z=0$, gas accretion is the dominant contributor to the BH mass assembly only for sufficiently massive $\gtrsim10^7~M_{\odot}$ BHs. But at the low mass end of $\sim10^5-10^6~M_{\odot}$, mergers continue to be the dominant mode of BH growth at $z=0$. This explains why the signatures of BH seeding remain strong within $\sim10^5-10^6~M_{\odot}$ all the way down to $z=0$. Furthermore, the transition from merger dominated growth at $\sim10^5-10^6~M_{\odot}$ to accretion dominated growth at $\gtrsim10^7~M_{\odot}$ leaves a curious imprint on the $z=0$ BHMFs. Specifically, the slope of the $z=0$ BHMFs is somewhat steeper at $\sim10^5-10^6~M_{\odot}$ compared to $\gtrsim10^7~M_{\odot}$. This transition in slope does not occur for the $z=5$ BHMFs wherein the entirety of BH growth is merger dominated. We also note that the commonly used Schechter function modeling of the $z=0$ BHMFs would not be able to capture the steepening of the slope at $\sim10^5-10^6~M_{\odot}$. To that end, if this transition in the $z=0$ BHMF slope indeed occurs in the real Universe, a naive extrapolation of the observed BHMFs at $\gtrsim10^6~M_{\odot}$ (which were indeed modeled based on Schechter functions) to lower masses, would underestimate the abundances of $\sim10^5-10^6~M_{\odot}$ BHs.  

To summarize our findings so far, the local $\sim10^5-10^6~M_{\odot}$ BHs have strong signatures of seed models. This is because of two reasons. First, there is a substantial population of the lowest mass $\sim10^5~M_{\odot}$ BHs at $z=0$ that are ungrown relics of high redshift seeds~($z\sim5-10$). Second, the BH growth from $\sim10^5$ to $\sim10^{6}~M_{\odot}$ is dominated by mergers all the down to $z\sim0$. This merger-driven growth tends to retain the  signatures of seed models in the abundances of $\sim10^5-10^6~M_{\odot}$ BHs. However, as accretion-driven growth becomes increasingly important for higher mass BHs, the signatures of seeding begin to be erased and eventually disappear for $\gtrsim10^7~M_{\odot}$ BHs.

During the writing of this paper, \cite{2024arXiv240906772T} published BH mass functions for broad line~(BL) AGNs at $z\sim4-6$ detected using JWST~(shown as black stars in right panel of Figure \ref{mass_function}). While the comparison against JWST observations was the focus of our previous paper~\citep{2024MNRAS.533.1907B}, it is instructive to take a detour and compare these recent results with our simulation predictions, before bringing back our focus on $z=0$. Our overall $z=5$ BHMFs is significantly higher than these observations~(black stars vs solid lines in Figure \ref{mass_function}). However, \cite{2024arXiv240906772T} only probes BHs that have luminosities above the detection limit. Notably, our simulated BHMFs of active BHs~(above $L_{\rm bol}>10^{43}~\rm erg~s^{-1}$, dashed lines in Figure \ref{mass_function}) is substantially smaller than the overall BHMFs, and is much closer to the observational constraints. All this is again tied to the fact that the BH mass assembly at high-z is merger dominated. In such a case, we could expect the existence of a much larger population of inactive merging BHs that cannot be probed by JWST. The JWST based measurements of \cite{2024arXiv240906772T} would then serve as a strict lower limit to the overall BHMFs at high-z. Future facilities such as LISA will be able to test for the possible existence of a larger population of inactive BHs that is present in our simulations.

\subsection{AGN luminosity functions}
\begin{figure*}
\includegraphics[width=17 cm]{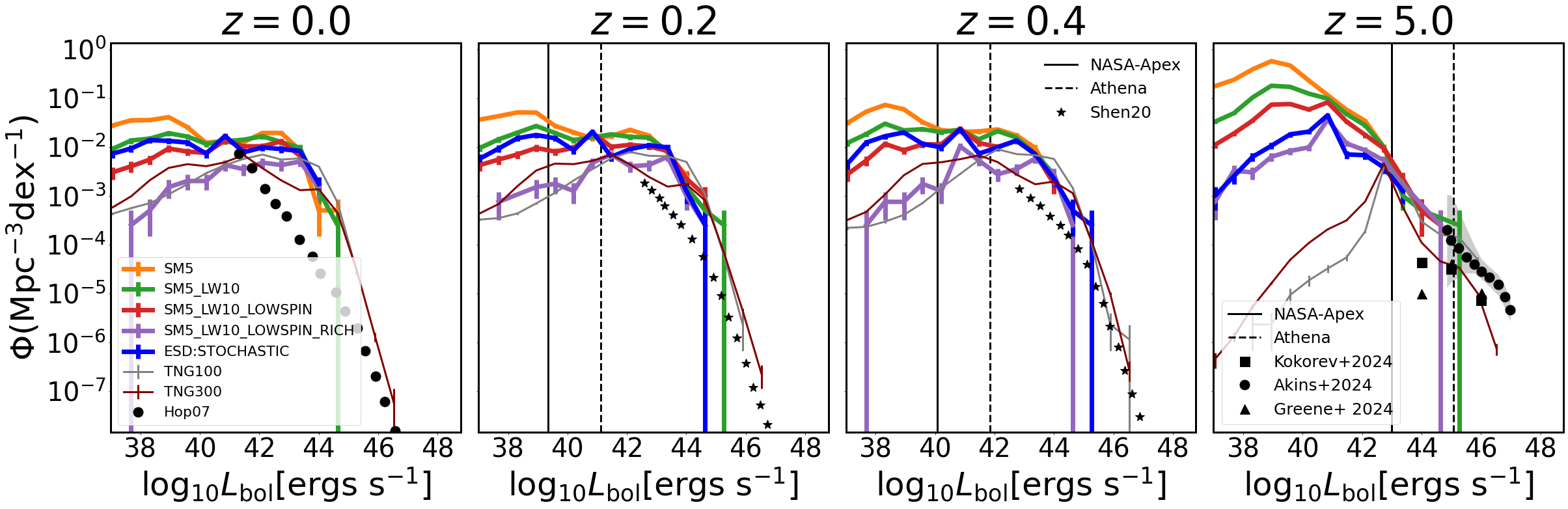}

\includegraphics[width=17 cm]{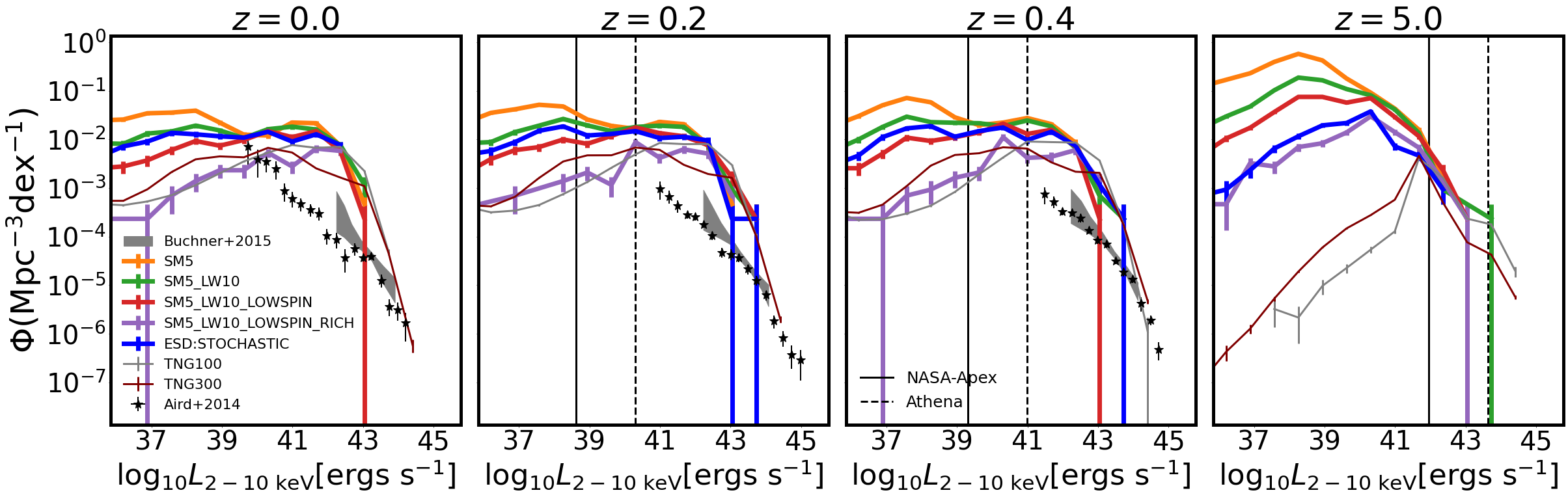}
\caption{The top and bottom rows show the AGN bolometric and hard X-ray luminosity functions respectively, for the different seed models. The left three columns show different redshifts in the nearby Universe whereas the rightmost column shows the $z=5$ prediction. The thin grey and maroon lines show predictions TNG100 and TNG300 respectively. The black points correspond to observational constraints from \protect\cite{2020MNRAS.495.3252S}, \protect\cite{2015MNRAS.451.1892A} and \protect\cite{2015ApJ...802...89B}. 
All the seed models converge at $L_{\rm2-10 ~keV}\gtrsim10^{42}~\mathrm{erg~s^{-1}}$ or $L_{\rm bol}\gtrsim10^{43}~\mathrm{erg~s^{-1}}$. The strongest seed model variations in the AGN LFs
are at $L_{\rm bol} \lesssim 10^{40}~\mathrm{erg~s^{-1}}$ and $L_{\rm 2-10~keV} \lesssim 10^{39}~\mathrm{erg~s^{-1}}$. With proposed X-ray facilities such as AXIS, these luminosities can be accessible in the local Universe, but not in the high-z Universe.}
\label{AGN_luminosity_functions_fig}
\end{figure*}

The 1st three columns of Figure \ref{AGN_luminosity_functions_fig} show the AGN luminosity functions~(LFs) close to the local Universe; i.e. $z\sim0-0.4$. We show both the bolometric luminosities as well as the X-ray luminosities. Note that while the simulations are able to readily predict the bolometric luminosities from the accretion rates~(modulo the uncertainties in the accretion and feedback model), they require a bolometric correction to infer the X-ray luminosities. Conversely, while the observations are able to directly measure the X-ray luminosities (barring uncertainties due to obscuration and X-ray binary contributions), they require a bolometric correction to infer the bolometric luminosities. Here we use the bolometric correction from \cite{2020MNRAS.495.3252S}. 

For sufficiently bright AGNs~($L_{\mathrm{bol}}>10^{43}~\mathrm{erg~s^{-1}}$ and $L_{\mathrm{2-10~keV}}>10^{42}~\mathrm{erg~s^{-1}}$), the LFs for all the different seed models converge. The \texttt{Illustris-TNG} seed model also produces similar LFs to the \texttt{BRAHMA} boxes in this regime~(see thin maroon and grey lines). The abundances of the brightest quasars ($L_{\mathrm{bol}}>10^{45}~\mathrm{erg~s^{-1}}$ or $L_{\mathrm{2-10~keV}}>10^{44}~\mathrm{erg~s^{-1}}$), probed only by the much larger TNG boxes, are broadly consistent with observations. However, for AGNs with $L_{\mathrm{bol}}\lesssim10^{44}~\mathrm{erg~s^{-1}}$, all the simulated LFs are greater than the observations by a factor of $\sim10$. This was also identified in previous works~\citep{2018MNRAS.479.4056W,2021MNRAS.507.2012B}. The same is true for several other simulations in the literature~(see Figure 5 of \citealt{2022MNRAS.509.3015H}). Recall from the previous section that the local BHMFs exhibit much better agreement between the simulations and observations~(particularly for $\gtrsim10^{7}~M_{\odot}$ BHs). This is because the local BHMFs are derived from BH-galaxy scaling relations and the galaxy luminosity functions, both of which are reasonably well reproduced by the Illustris-TNG galaxy formation model upon which \texttt{BRAHMA} is based~\citep{2020MNRAS.492.5167V,2020MNRAS.493.1888T}. The agreement in the BHMFs between the simulations vs. observations hints that at least part of the discrepancy in AGN LFs may be contributed by AGN obscuration. The uncertainties in the bolometric corrections as well as modeling uncertainties in the accretion prescription can also play a significant role. In any case, our seeding prescriptions are not consequential to this discrepancy. Therefore, we defer a detailed investigation of this discrepancy to a future paper. 

Despite the current tension with the observed $z=0$ LFs, the simulation predictions for the different seed models help us identify AGN populations wherein there are strong signatures of seeding. The seed model variations in AGN LFs become strong at $L_{\rm bol}\lesssim10^{40}~\mathrm{erg~s^{-1}}$ and $L_{\rm 2-10~keV}\lesssim10^{39}~\mathrm{erg~s^{-1}}$. For example, the differences between the LF predictions for \texttt{SM5} and \texttt{SM5_LW10_LOWPSIN_RICH} boxes become larger than factors of $\sim100$ at these faint luminosities. While these luminosities are substantially below what the current observations probe, they may be accessible by proposed X-ray facilities such as Athena~\citep{2020AN....341..224B} and the Advanced X-ray Imaging Satellite or AXIS~\citep{2023SPIE12678E..1ER}. In particular, AXIS could potentially detect AGNs all the way down to a few times $\sim10^{38}~\mathrm{erg~s^{-1}}$ in the local Universe. Below those luminosities, X-ray binaries are expected to become a substantial source of contamination~\citep{gallo2023blackholeoccupationfraction}. 

Finally, we compare our local AGN LFs to that at $z=5$ shown in the rightmost column of Figure \ref{AGN_luminosity_functions_fig}. The seed model variations at $z=5$ exhibit a similar trend to that at $z=0$; i.e. smaller variations at the brightest end~($L_{\rm 2-10~ keV}\gtrsim10^{43}~\mathrm{erg~s^{-1}}$) and larger variations at the faint end~($L_{\rm 2-10~keV}\lesssim10^{39}~\mathrm{erg~s^{-1}}$). Therefore, both $z=5$ and $z=0$ LFs have  a similar luminosity range of $L_{\rm 2-10~keV}\lesssim10^{39}~\mathrm{erg~s^{-1}}$ wherein we expect signatures of seeding. However, these luminosities are going to be inaccessible at $z\sim5$ even with future X-ray facilities. The recent JWST observations of high-z AGN~(black points in Figure \ref{AGN_luminosity_functions_fig}) also only probe down to $L_{\rm bol}\sim10^{44}~\mathrm{erg~s^{-1}}$ at $z\sim4-6$ wherein the LFs are similar for all the seed models. As discussed in our previous paper~\citep{2024MNRAS.533.1907B}, our simulations are in broad agreement with the JWST LFs at $z\sim5$~(albeit the observations currently have substantial uncertainties). 

We see yet another striking difference in the seed model variations amongst the \textit{overall} BH vs. the \textit{active} AGN populations at $z=0$ and $z=5$. At $z=0$, the seed model variations in the AGN LFs showed a similar trend to the BHMFs; i.e. seeding signatures become weaker for more massive BHs as well as more luminous AGNs at $z=0$. But at $z=5$, the abundances of the most luminous AGNs have no seed model variations whereas that of the most massive BHs have the strongest seed model variations. The reason for all this goes back to the mergers vs. accretion dominated BH growth at various redshifts shown in Figure \ref{mergers_vs_accretion}. At $z=5$, the BHMFs are ``de-coupled" from the "AGN LFs" due to the merger-dominated BH growth. In such a case, only a select set of BHs end up in environments with enough fuel to become luminous AGN regardless of the seed model. However, the formation of massive BHs at $z=5$ relies only on the availability of sufficient seeds. In contrast, close to $z=0$, gas accretion predominantly drives the growth of massive $\gtrsim10^7~M_{\odot}$ BHs. Therefore, at $z=0$, the BHMFs are ``coupled" to the AGN LFs, such that the formation of massive BHs essentially depends on the availability of accreting gas rather than the number of seeds.             

Overall, while we concluded in previous works~\citep{2021MNRAS.507.2012B,2024MNRAS.529.3768B,2024MNRAS.533.1907B} that AGN observations at higher-z~($z\sim5-10$) may not be very informative about seed models, the same is not true in the local Universe. Somewhat counter-intuitively, the local AGN LFs have a better prospect of constraining seed models compared to the higher-z AGN LFs. But this is simply because in the local Universe, future X-ray missions have the ability to probe exceptionally low X-ray luminosities~(down to $\sim 10^{38}~\mathrm{erg~s^{-1}}$) wherein our simulations show strong signatures of seeding.



\subsection{Stellar mass vs BH mass scaling relations}

\begin{figure*}
\includegraphics[width=18 cm]{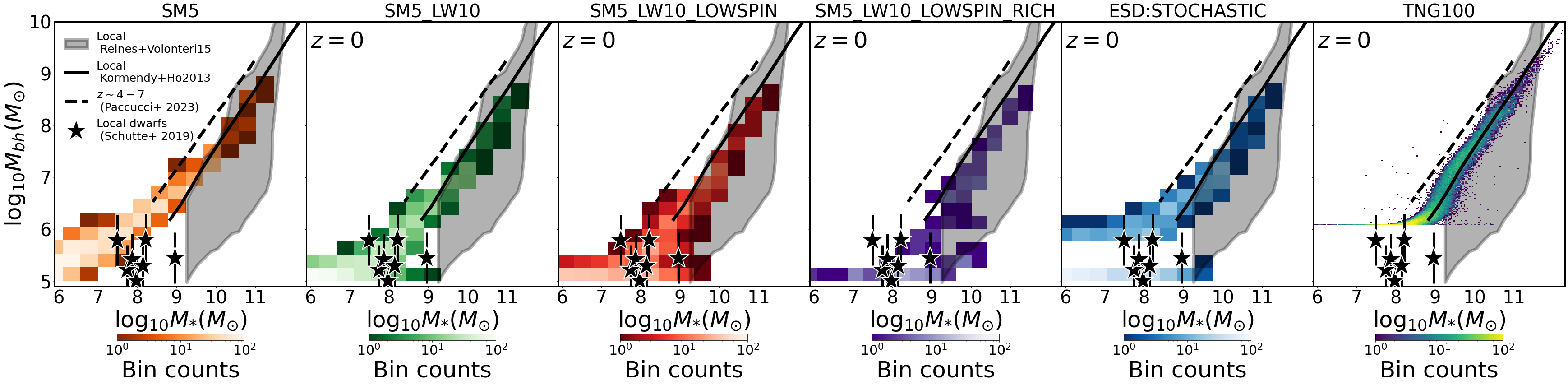}
\includegraphics[width=18 cm]{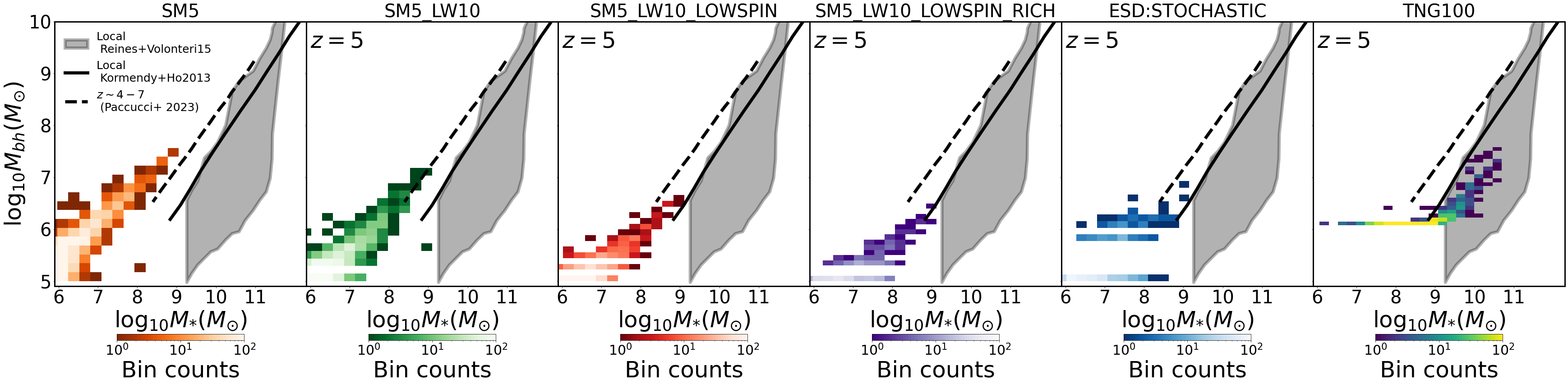}
\caption{$M_{*}$ vs. $M_{bh}$ relation predictions for our different BH seed models compared against different observational measurements. Specifically, we plot the total stellar mass of the subhalos versus the mass of central~(most massive) BH. The four left panels show the DHS models. The fifth panels shows the ESD model. The rightmost panels show the results from TNG100. The top rows show the local scaling relations compared against observations for $\gtrsim10^6~M_{\odot}$ BHs from \protect\citealt{2013ARA&A..51..511K}~(black solid line) and \protect\citealt{2015ApJ...813...82R}~(shaded region), as well as $\sim10^5-10^6~M_{\odot}$ BHs within dwarf galaxies from \protect\citealt{2019ApJ...887..245S}~(black stars). The bottom row shows the high-z~($z=5$) scaling relation compared against the relation derived by \protect\citealt{2023ApJ...957L...3P}~(black dashed line) for JWST AGNs at $z\sim4-7$. The different columns show the different seed models. For local galaxies with $M_*\gtrsim10^{10}~M_{\odot}$, all the seed models produce similar $M_{*}$ vs. $M_{bh}$ relations that are also consistent with observations. However, for local dwarf galaxies~($M_*\lesssim10^{9}~M_{\odot}$), the $M_{*}$ vs. $M_{bh}$ relations have significant seed model variations. Compared to the \protect\cite{2019ApJ...887..245S} measurments, the most optimistic DHS model~(\texttt{SM5}) predictions are higher, while the reamaining seed models are broadly consistent.} 
\label{scaling_relations}
\end{figure*}

We now focus on the relation between the BHs and their host galaxies at $z=0$. In the top panels of  Figure \ref{scaling_relations}, we show the local stellar mass vs. BH mass~($M_*$ vs $M_{bh}$) relations for the different seed models, compared against the observations. We note that our simulations are not large enough to robustly probe the scatter of the $M_*$ vs. $M_{bh}$ relations~(particularly in higher mass galaxies) as revealed in the observational measurements. This is because in these volumes, it is difficult to form (rarer) BHs that are significantly further away from the mean relation. Nevertheless, the simulations do capture the mean $M_*$ vs. $M_{bh}$ relations for the different seed models, which we can compare against observations. For $\gtrsim10^9~M_{\odot}$ galaxies, the $M_*$ vs. $M_{bh}$ relations are similar for all the seed models; these predictions are also in good agreement with measurements from \cite{2013ARA&A..51..511K} and \cite{2015ApJ...813...82R}. Notably, these galaxies host $\gtrsim10^7~M_{\odot}$ BHs wherein the majority of the mass growth occurs via gas accretion. 

For dwarf galaxies~($\lesssim10^9~M_{\odot}$) that typically host BHs between $\sim10^5-10^7~M_{\odot}$ growing largely via mergers, the $M_*$ vs. $M_{bh}$ relations exhibit non-negligible seed model variations. For the most lenient DHS model~(\texttt{SM5}), the slope of the scaling relations in $M_*\lesssim10^9~M_{\odot}$ galaxies becomes slightly shallower than that in $M_*\gtrsim10^9~M_{\odot}$ galaxies. This is due to the strong merger-driven growth of BHs in $M_*\lesssim10^9~M_{\odot}$ galaxies, fueled by the greater availability of seeds. For the more restrictive DHS models, there are fewer seeds to fuel merger-driven growth in $M_*\lesssim10^9~M_{\odot}$ galaxies. This leads to smaller BH masses~(at fixed stellar mass) in $M_*\lesssim10^9~M_{\odot}$ galaxies compared to the more lenient DHS models. Therefore, for the \texttt{SM_LW10_LOWSPIN} and \texttt{SM_LW10_LOWSPIN_RICH} DHS models, the change in slope of the scaling relations between $M_*\gtrsim10^9~M_{\odot}$ and $M_*\lesssim10^9~M_{\odot}$ is less prominent compared to the \texttt{SM5} model. Lastly, the ESD model shows a somewhat larger scatter in the $M_*$ vs $M_{bh}$ relations for $M_*\lesssim10^9~M_{\odot}$ galaxies, compared to the DHS models. This may be because in the ESD seed model, the $\sim10^5~M_{\odot}$ BHs are descendants of unresolved lower mass seeds that presumably went through a diverse range of (merger-driven) growth histories; we shall investigate this further in the future using larger volume simulations that will be able to quantify the scatter more robustly. 

We can now compare our predictions in the dwarf galaxy regime to observations from \citealt[][(black stars in Figure \ref{scaling_relations})]{2019ApJ...887..245S}. The most lenient \texttt{SM5} model generally predicts $\sim5-10$ times higher BH masses in $M_*\lesssim10^9~M_{\odot}$ galaxies compared to the observations. The remaining seed models are broadly consistent with the observations. These results demonstrate that in the future when the sample sizes are larger and measurement uncertainties are smaller, we might be able to discriminate between seeding scenarios based on the $M_*$ vs. $M_{bh}$ relations of BHs in local dwarf galaxies. However, to be able to do so, we also need to understand the impact of BH accretion, dynamics and feedback modeling in this regime; we will address this in future work.

The bottom panels of Figure \ref{scaling_relations} shows the $M_*$ vs $M_{bh}$ relations at $z=5$~(which was the focus of our previous paper ~\citealt{2024MNRAS.533.1907B}). In contrast to $z=0$ wherein the seed models impact the $M_*$ vs. $M_{bh}$ relations only in the dwarf galaxy regime, the $z=5$ relations are significantly impacted even in the most massive galaxies~(and BHs) probed by our simulation volumes~(i.e $M_*\sim10^{10}~M_{\odot}$). This again, is because mass growth is dominated by mergers for all the BHs at $z=5$~(and higher redshifts). Based on recent observations of JWST, \cite{2023ApJ...957L...3P} inferred an intrinsic $M_*$ vs. $M_{bh}$ relation~(grey solid line) that is overmassive compared to the local scaling relations by $>3\sigma$. Our intermediary DHS models~(\texttt{SM5_LW10} and \texttt{SM5_LW10_LOWSPIN}) and the ESD model produce the best agreement with the inferred high-z relations. It is noteworthy that these are also the same models which produce the best agreement with the most recent observational estimates of the local BH mass functions at $M_{bh}\lesssim10^7~M_{\odot}$~\citep[][revisit Figure \ref{mass_function}]{2020MNRAS.495.3252S} as well as the observed $M_{bh}$ to $M_*$ ratios for the local dwarf galaxies~(top panels of Figure \ref{scaling_relations}). On the other hand, the \texttt{SM5} model that overestimates the high-z $M_*$ vs. $M_{bh}$ relations, also slightly overestimates the local BH mass functions at $M_{bh}\lesssim10^7~M_{\odot}$~(and also the $M_{bh}$ to $M_*$ relations for local dwarf galaxies). Likewise, the \texttt{SM5_LW10_LOWSPIN_SPIN} model that underestimates the high-z $M_*$ vs. $M_{bh}$ relations, also tends to be at the lower end of the different observed local BHMFs at $M_{bh}\lesssim10^7~M_{\odot}$. In other words, the seeding models that have the best agreement with the inferred high-z overmassive $M_*-M_{bh}$ relation using JWST, also produce good agreement with the existing local observations. While the current observational uncertainties are relatively high, it is encouraging to see that three of our seed models are simultaneously consistent with current observations of BHs at the high-z as well as the local Universe. 

Lastly, we can also compare the \texttt{BRAHMA} predictions to that of \texttt{TNG100} shown in the rightmost panels of Figure \ref{scaling_relations}. We can readily notice that unlike the \texttt{BRAHMA} boxes, the \texttt{TNG100} scaling relations flatten~(zero slope) at their adopted seed mass of $\sim10^6~M_{\odot}$ in the dwarf galaxy regime of $M_*\lesssim10^{9}~M_{\odot}$. Essentially, this is because there are not enough seeds produced in TNG100 to fuel merger-driven BH growth within $M_*\lesssim10^{9}~M_{\odot}$ galaxies wherein gas accretion is suppressed by stellar feedback. Concurrently, due to the lack of merger driven BH growth, the \texttt{TNG100} scaling relations help us easily visualize the fact that at both $z=0~\&~5$, accretion-driven BH growth starts to be significant only when the stellar mass exceeds $M_*\gtrsim10^9~M_{\odot}$. This explains why our \texttt{BRAHMA} boxes do not see substantial accretion-driven BH growth at $z=5$, as they are too small to produce $M_*\gtrsim10^9~M_{\odot}$ galaxies at these redshifts. At $z=0$ however, $M_*\gtrsim10^9~M_{\odot}$ galaxies become common enough to be captured by our \texttt{BRAHMA} boxes, wherein we readily see accretion-driven BH growth.

\subsection{BH occupation fractions}
\label{BH occupation fractions sec}
\begin{figure*}
\includegraphics[width=18 cm]{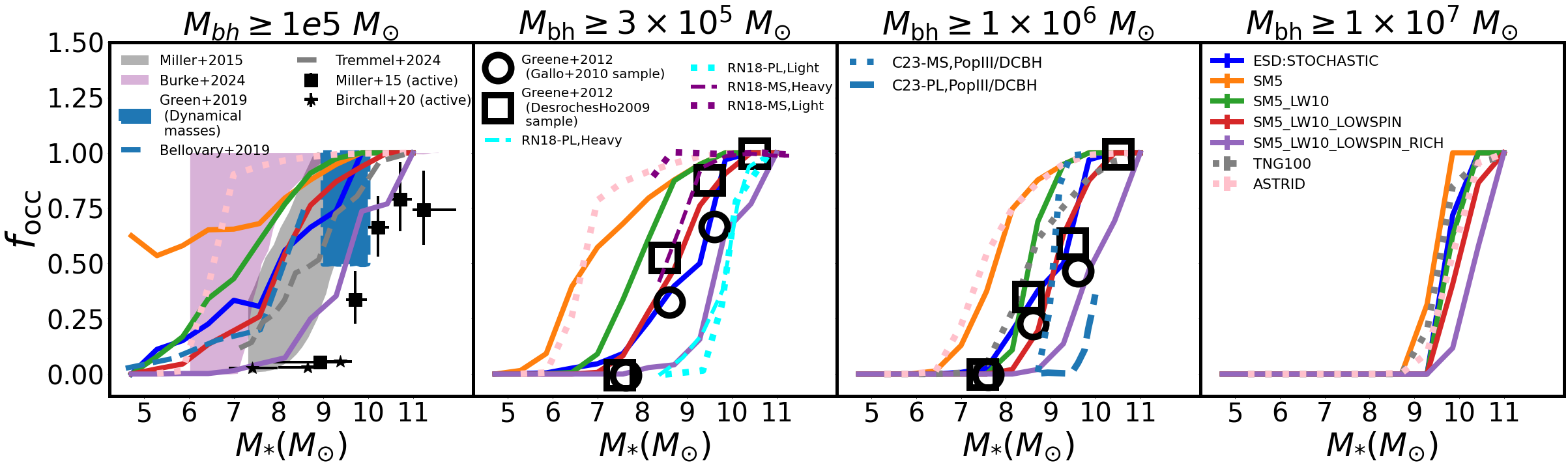}
\caption{$z=0$ occupation fractions predicted by the different seed models. The different panels correspond to BH samples with different mass thresholds. Seed model variations are substantial for $\gtrsim10^5~M_{\odot}$ and $\gtrsim10^6~M_{\odot}$ BHs. We compare our predictions with observational constraints as well as several other theoretical models. In the 1st panel, dashed blue and grey lines  are hydrodynamic simulations of \protect\cite{2019MNRAS.482.2913B} and \protect\cite{2024OJAp....7E..26T}, respectively. The black squares are observational measurements of the active BH occupation fractions from \protect\cite{2015ApJ...799...98M}, along with the grey region being their inferred overall occupation fraction. The blue region derived by \protect\cite{2020ARA&A..58..257G} corresponds to the only constraints that are derived from dynamical mass measurements of a sample of 10 BHs in \protect\cite{2019ApJ...872..104N}. Purple and cyan dashed and dotted lines in the 2nd panel are predictions from the semi-analytical model of  \protect\citealt{2018MNRAS.481.3278R}~(R18). In the 3rd panel, the blue dashed and dotted lines are predictions from \protect\cite{2023ApJ...946...51C} using the semi-analytic model of R18. The black squares and circles are the early measurements from \protect\cite{2012NatCo...3.1304G} based on samples of \protect\cite{2009ApJ...698.1515D} and \protect\cite{2010ApJ...714...25G} respectively. The purple shaded region corresponds to the most recent constraints for the overall occupation fraction by \protect\cite{burke2024multiwavelengthconstraintslocalblack}. The grey and pink dotted lines are predictions from \texttt{TNG100} and \texttt{ASTRID} respectively. Our predictions are broadly consistent with observations, with current uncertainties being too large to discriminate between seeding scenarios.}

\label{BH_occupation_fractions}
\end{figure*}

The BH occupation fractions in local dwarf galaxies is regarded as one of the most promising observables to constrain seed models in the local Universe. 
The local BH occupations for our different simulations are shown in Figure \ref{BH_occupation_fractions} for different BH mass thresholds. The leftmost panel shows the smallest threshold we can probe in our simulations; i.e. $\gtrsim10^5~M_{\odot}$. We find that the occupation fractions for $\gtrsim10^5~M_{\odot}$ BHs are  substantially different between the different seed models for $M_*\lesssim10^9~M_{\odot}$ galaxies. For $\sim10^9~M_{\odot}$ galaxies, the most lenient DHS model predicts an occupation fraction of $\sim100\%$ whereas the most restrictive DHS model predicts $\sim40\%$ occupation fractions. The ESD model predicts occupation fractions $\sim70~\%$ in $\sim10^9~M_{\odot}$ galaxies. The seed model variations are stronger for even lower mass galaxies. 

Much like the BH mass functions, the seed model variations of the BH occupation fractions also tend to become smaller for increasing BH masses. However, even for $\gtrsim10^6~M_{\odot}$ BHs, the BH occupation fractions have significant seed model variations. For the $M_*\sim10^9~M_{\odot}$ galaxies, the BH occupation fractions vary from $20~\%$ to $100~\%$ between the most restrictive to the most lenient DHS model. For even higher mass thresholds of $M_{bh}\gtrsim10^7~M_{\odot}$, all of our seed models predict broadly similar results. 
\subsubsection{Comparison with observations}
Several attempts have been made to observationally constrain the BH occupation fractions. Some of the early measurements were made by \cite{2012NatCo...3.1304G} using X-ray AGN samples from \cite{2009ApJ...698.1515D} and \cite{2010ApJ...714...25G}. However, since these measurements only target \textit{active} AGNs, one needs to make assumptions for the AGN fractions~(fraction of BHs that are active and detectable) in order to infer the occupation fractions of the full BH population~(active as well as inactive). By assuming $10~\%$ detectable AGN fractions, and counting only BHs that are above Eddington ratios of $10^{-4}$, \cite{2012NatCo...3.1304G} derived the occupations fractions of $>3\times10^5~M_{\odot}$ and $>10^6~M_{\odot}$ BHs~(black circles and squares in 2nd panel of Figure \ref{BH_occupation_fractions}). A few years later, \cite{2015ApJ...799...98M} used a similar procedure on a much larger sample of X-ray AGNs and inferred the active as well as the full BH occupation fractions~(black stars and grey region in the 1st panel). Since our simulated AGN LFs are currently higher than the observed LFs at the faint end~(revisit Figure \ref{AGN_luminosity_functions_fig}), it is difficult to replicate the selection criteria used in these works and perform an even-handed comparison between simulated and observed occupation fractions. Nevertheless, at the very least, it is encouraging to see that the \cite{2012NatCo...3.1304G} and \cite{2015ApJ...799...98M} results broadly fall within the range of our simulation predictions for the different seed models~(2nd and 3rd panel of Figure \ref{BH_occupation_fractions}). Additionally, the active occupation fractions from \citealt{2015ApJ...799...98M}~(black squares) and \citealt{2020MNRAS.492.2268B}~(black stars) serve as lower limits to the overall BH occupation~(1st panel of Figure \ref{BH_occupation_fractions}); they also do not rule out any of our seed models since all of the simulations predict higher overall BH occupations compared to their results~(see 1st panel of Figure \ref{BH_occupation_fractions}). Very recently, \cite{burke2024multiwavelengthconstraintslocalblack} derived multi-wavelength constraints for the overall BH occupation fraction by combining X-ray, radio and optical variability data. Their lower limits for the occupation fractions~(pink shaded region in leftmost panels) are higher than our two restrictive DHS models. However, they are not in conflict with the predictions of our optimistic DHS models. 
The \cite{burke2024multiwavelengthconstraintslocalblack} constraints are also higher than our ESD model predictions; this is not surprising because their constraints do not assume a specific BH mass threshold, whereas we do not explicitly resolve the low mass seeds that the ESD model attempts to represent.

The most even-handed comparison for our $M_{bh}>10^5~M_{\odot}$ occupation fractions would be against the results from \cite{2020ARA&A..58..257G}, as they used dynamical BH mass measurements~\citep{2019ApJ...872..104N} to estimate the occupation fraction of $M_{bh}\gtrsim10^5~M_{\odot}$ BHs in a sample of $\sim10^9-10^{10}~M_{\odot}$~(albeit only for 10 galaxies). Based on BH detections in at least 5 out of 10 galaxies, \cite{2020ARA&A..58..257G} inferred $>50\%$ occupation fractions for $\gtrsim10^9$ galaxies; this is also consistent with all of our seed model predictions~(see blue region in the 1st panel of Figure \ref{BH_occupation_fractions}). In summary, the current measurements for the BH occupation fractions have too large uncertainties to strongly constrain our seed models. Future measurements with deeper X-ray probes like Athena and AXIS~(see \citealt{gallo2023blackholeoccupationfraction}), along with large sample of dynamically measured BH masses in dwarf galaxies, are going to be crucial for constraining seed models using local BH occupation fractions.  

\subsubsection{Comparison with other model predictions}

We also compare our predicted BH occupation fractions against several other simulations and semi-analytic models. We start with cosmological hydrodynamic simulations. \cite{2019MNRAS.482.2913B} and \citealt{2024OJAp....7E..26T}~(\texttt{ROMULUS}) use a similar seed model originally developed in \cite{2017MNRAS.470.1121T} wherein heavy $\sim10^5~M_{\odot}$ seeds are generated from individual gas cells that are high density~($\sim15$ times the SF threshold), low metallicity~($\lesssim3\times10^{-4}$), with temperatures between $9500-10000~\mathrm{K}$. Their occupations of $\geq10^5~M_{\odot}$ BHs~(leftmost panel of Figure \ref{BH_occupation_fractions}) are smaller than our most optimistic DHS model~(\texttt{SM5}) likely because they adopt a much higher gas density threshold for seed formation compared to ours. However, due to the addition of LW flux, gas spin and halo environment criteria, our most pessimistic DHS model~(\texttt{SM5_LW10_LOWSPIN_RICH}) predicts BH occupations smaller than these works. The intermediary DHS models~(\texttt{SM5_LW10} and \texttt{SM5_LW10_LOWSPIN}) and the ESD models produce BH occupation fractions broadly similar to \cite{2019MNRAS.482.2913B}. 

The \texttt{ASTRID}~\citep{2022MNRAS.513..670N,ni2024astridsimulationevolutionblack} simulation initializes BHs with a power law distribution of seed masses from $4.4\times10^4-4.4\times10^5~M_{\odot}$ in halos exceeding total masses of $7.4\times10^9~M_{\odot}$ and stellar masses of $3\times10^6~M_{\odot}$. The simulation has been so far run to $z=0.2$ with galaxy catalogs processed down to $z=0.4$~(Chen et al in prep). While it is yet to reach $z=0$, it is nevertheless insightful to compare the $z=0.4$ occupation fractions against our local results~(using data obtained by private communication). For $\gtrsim10^5~M_{\odot}$ BHs, the \texttt{ASTRID} occupation fractions are higher than all of our seed models within $M_*\gtrsim3\times10^6~M_{\odot}$ galaxies. However, due to the adopted stellar mass threshold, the occupation fractions sharply fall off in galaxies with $M_*\lesssim3\times10^6~M_{\odot}$. For $\gtrsim10^6~M_{\odot}$ BHs, the \texttt{ASTRID} occupation fractions are similar to our most optimistic \texttt{SM5} seed model.

The \texttt{IllustrisTNG} simulations seed $\sim10^{6}~M_{\odot}$ BHs in $7\times10^{10}~M_{\odot}$ halos. The relatively high halo mass threshold makes the TNG seed model substantially more restrictive than the \texttt{SM5} model. Therefore, despite the lower seed mass than \texttt{BRAHMA}, TNG predicts smaller occupation fractions for $\geq10^6~M_{\odot}$ BHs compared to the \texttt{SM5} model. However, the TNG predictions are higher than the \texttt{SM5_LW10_LOWSPIN_RICH} model. For the highest mass $\geq10^7~M_{\odot}$ BHs, the \texttt{BRAHMA}, \texttt{IllustrisTNG} as well as the \texttt{ASTRID} simulations predict broadly similar occupation fractions. This adds further support to our finding that for $\gtrsim10^7~M_{\odot}$ BHs, the signatures of seeding are largely erased. 

A general takeaway from the above results is that the BH occupation fractions predicted by \texttt{BRAHMA} span a broad range, within which the predictions from many hydrodynamic simulations fall comfortably. This serves as a demonstration of the diversity of the seed models explored within the \texttt{BRAHMA} simulations, some of which are significantly more lenient compared to other simulations in the literature, while others being significantly more restrictive. 

The semi-analytic model of \citealt{2018MNRAS.481.3278R} (hereafter R18) has been used to predict BH occupation fractions for different seeding and growth scenarios (R18; \citealt{2023ApJ...946...51C}). These are shown in the second and third panels of Figure \ref{BH_occupation_fractions}. These papers explored a ``Heavy" seed model with a seed mass distribution that peaks at $\sim10^5~M_{\odot}$), using the seeding criteria from \cite{2006MNRAS.371.1813L}~(which is essentially our \textit{gas-spin criterion}). They also explore a ``light" seed model which assumes a power-law seed mass distribution from $30-100~M_{\odot}$. R18 finds that the occupation fractions for $\geq3\times10^5~M_{\odot}$ and $\geq10^6~M_{\odot}$ BHs are also sensitive to the accretion scenario. Therefore, in order to interpret the differences between R18 vs. \texttt{BRAHMA}, we have to also account for the differences in the BH accretion models. R18 has two distinct modes of accretion, an episodic ``burst" mode and a "steady" mode. At $z\gtrsim2$, their accretion is mainly dominated by the ``burst mode", wherein for every major merger (merging halo mass ratios $>0.1$) the BHs are grown at Eddington until they reach a cap that corresponds to the local $M-\sigma$ relation. As already seen in Figure \ref{mergers_vs_accretion}, the \texttt{BRAHMA} simulations do not see any significant contribution from BH accretion on to the BH growth at $z\gtrsim3$. This essentially means that the episodic burst mode assumed by R18 is absent in \texttt{BRAHMA}. In the ``steady" mode accretion that is more significant at $z\lesssim2$, RN18 adopts a ``main sequence"~(MS) accretion model wherein the BH accretion rates are proportional to~(a thousandth) of the star formation rates. This mode allows for significantly faster growth of BHs in low mass halos compared to the \texttt{BRAHMA} simulations. Additionally, they also adopt a ``power law"~(PL) model wherein the Eddington ratios are drawn from a power-law distribution. This model exhibits slower growth of BHs in low mass halos, consistent with \texttt{BRAHMA}. All that being said, the most even-handed comparisons with the \texttt{BRAHMA} seed models would be with those R18 seed models that are coupled with the PL accretion model. We can see that for all the PL models~(cyan dashed lines in Figure \ref{BH_occupation_fractions}), the R18 occupation fractions are smaller than all the \texttt{BRAHMA} boxes. This is likely because R18 allows seed formation to occur only between $z\sim15-20$, whereas the seed formation rates in \texttt{BRAHMA} are suppressed only after $z\sim10$. Since the MS models of R18 allow for faster BH accretion in low mass halos, they predict higher occupation fractions compared to the PL models. As a result, the MS model predictions of R18 lie within the span of our \texttt{BRAHMA} predictions despite their seed models being more restrictive. 

To summarize, we find that the BH occupation fractions in local dwarf galaxies contain strong imprints of seeding, with predicted seed model variations for $\gtrsim10^5~M_{\odot}$ and $\gtrsim10^6~M_{\odot}$ BHs ranging from $\sim40-100~\%$ and $\sim20-100~\%$ respectively for $M_*\sim10^9~M_{\odot}$ galaxies. However, other studies such as R18 and \cite{2023ApJ...946...51C} have shown that they may also be sensitive to the BH accretion models in addition to our seeding models. In the future, we plan to explore the dependence of BH occupation fractions on different choices for accretion models.

\section{Discussion}
\label{Discussion sec}
\subsection{Feasibility of low mass seeds and heavy seeds as origins of the local SMBH populations}

We now use our results to discuss the feasibility of low mass seeds vs. heavy seeds in explaining the local SMBH populations. When we look at the different local BH observables in Figures \ref{Number_density} to \ref{BH_occupation_fractions}, we find that the ESD model predictions are in good broad agreement with their observational counterparts, particularly for the abundances of $10^6-10^7~M_{\odot}$ and the $M_*-M_{bh}$ relations in dwarf galaxies. This suggests that the lower mass $\sim10^3~M_{\odot}$ seeds may be viable origins for the local SMBH populations. While our ESD model is agnostic to which physical low mass seeding scenarios it may represent, it is worth noting that the target seed masses of $\sim10^3~M_{\odot}$ best resemble NSC seeds or extremely massive Pop III seeds. 

As for the DHS models, there is a significant spread in predictions at the lowest mass end~$\sim10^5-10^6~M_{\odot}$ that is primarily composed of local ungrown relics of seeds that form at high redshift~($z\sim5-10$). The most lenient \texttt{SM5} model predicts number densities and BHMFs slightly above the upper end of the observational measurements. While the BHMF estimates can be impacted by uncertainties in the BH-galaxy scaling relations, the \texttt{SM5} model also overpredicts the $M_{bh}/M_{*}$ ratios in the dwarf galaxy regime compared to measurements of \cite{2019ApJ...887..245S}. All this is not surprising because \texttt{SM5} assumes that every halo with sufficient dense and metal-poor gas will form DCBHs; this is likely to be too optimistic as we expect the need for additional conditions to prevent the cooling and fragmentation with the required dissociation of molecular Hydrogen. When we add the \textit{LW flux criterion} and the \textit{gas spin criterion}, the resulting BH number densities, BHMFs and $M_{bh}/M_{*}$ ratios are broadly consistent with observations. However, the relatively low value of $10~J_{21}$ for the critical LW flux may only be feasible when there is additional help from dynamical heating~(in major halo mergers \citealt{2019Natur.566...85W,2020OJAp....3E...9R}) in the fighting against $H_2$ cooling and fragmentation. For this reason, we added an additional criterion in \texttt{SM5_LW10_LOWSPIN_RICH} that requires DCBH formation to occur only in rich environments with neighboring massive halos. For this most restrictive DHS model, the predicted BH number densities and BHMFs are already at the lower end of the observed measurements~(revisit Figures \ref{Number_density} and \ref{mass_function}); the predicted BH occupation fractions are also below the latest lower limits derived by \cite{burke2024multiwavelengthconstraintslocalblack}. These results imply that any DCBH formation mechanism that is substantially more restrictive than our \texttt{SM5_LW10_LOWSPIN_RICH} model, would not be viable for explaining the bulk of our local SMBHs.  

To that end, DCBH formation has long been considered to be possible only under extremely restrictive and therefore rare environments, requiring LW fluxes above $1000~J_{21}$~\citep{2010MNRAS.402.1249S,2014MNRAS.445..544S,2017MNRAS.469.3329W}. These are much more stringent than any of our DHS models. In fact, in \cite{2022MNRAS.510..177B}, we showed that these flux thresholds would produce only a handful of BHs~(if any) in our \texttt{BRAHMA} volumes. In such a case, lower mass seeds~(as in our ESD seed model) would be the only viable origins for the bulk of the local BH populations. However, we are far from having a complete theoretical understanding of the full details of heavy seed formation. In fact, the recent discovery of potentially overmassive BHs by JWST~(see section \ref{introduction}) also suggest the possibility that heavy seeds form more frequently than previously thought. Therefore, if we remain agnostic about the formation efficiency of heavy seeds, both low mass and heavy seeds can be viable origins for our local SMBHs. This would make it challenging to ascertain whether our local SMBHs originated from low mass seeds vs heavy seeds~(or both). 

The above arguments also suggest that knowing the formation efficiency of heavy seeds is going to be crucial for determining the seeding orgins of local SMBH populations. In order to probe the formation efficiency of heavy seeds, we might have to rely on the observations of $\sim10^5~M_{\odot}$ BHs at high redshifts~($z\gtrsim6$), wherein there is a clear distinction in the number density evolution of $\gtrsim10^5~M_{\odot}$ BHs between the DHS models and the ESD model. In particular, all the DHS models predict much higher number densities than the ESD model at $z\gtrsim6$~(revisit left panel of Figure \ref{Number_density}). 
This may have significant consequences on the predicted rates of GW events from merging $\gtrsim10^5~M_{\odot}$ BHs between low mass vs heavy DCBH seed models at $z\gtrsim6$. We explore this in ongoing work~(Bhowmick et al in prep).

\subsection{Caveats: Uncertainties in modeling of BH accretion, feedback and dynamics}
\label{caveats}

One of the big forthcoming challenges in learning about BH seeding from current and future observations, is to navigate potential degeneracies due to uncertainties in the modeling of several other physical processes such as BH accretion, dynamics, star formation, metal enrichment and feedback. To begin with, since our seed models depend on the formation of dense and metal poor gas, alternative treatments for star formation and metal enrichment can significantly impact the rates of seed formation at different redshifts. The subsequent growth of these seeds via accretion and mergers can be impacted by the modeling of BH accretion and dynamics. To that end, our work identifies strong signatures of seeding in regimes where accretion is suppressed by stellar feedback and the BH growth is dominated by mergers. However, this may be sensitive not just to the choice of the BH accretion model, but also the stellar feedback model.

Notably, when stellar feedback is sufficiently strong, BH accretion may be substantially suppressed regardless of the choice for our accretion model. Recent work by Burger et al~(in prep) suggests that this may indeed be the case for the TNG stellar feedback model inherited by our simulations. In such a case, the merger dominated BH growth in low mass galaxies may be fairly robust to the choice of the BH accretion model. However, if the stellar feedback is not very strong, the BH growth in low mass galaxies could become much more sensitive to the accretion model. In such a case, the implementation of AGN feedback can also have a profound impact on BH growth. While AGN feedback is  well-accepted to be important in massive galaxies, recent observations have also shown evidence for AGN-driven gas outflows in dwarf galaxies~\citep{2019ApJ...884...54M}. In our AGN feedback model adopted from TNG, we would not produce outflows from low mass BHs as the ``kinetic mode" does not switch on until the BH masses are above $\sim10^8~M_{\odot}$~(and the ``thermal mode" feedback is generally ineffective in producing outflows). This suggests the need for exploring alternative feedback implementations that can allow for the AGN kinetic feedback to operate in galaxies with low mass BHs~\citep[for e.g.][]{2024arXiv240902172K}.


Another major source of uncertainty is the BH dynamics modeling. In this work, by repositioning the BHs, we assume the most optimistic scenario for merger-driven BH growth. In reality, the merging efficiency will depend on the time scales associated with the hardening of BH binaries due to processes like dynamical friction,  stellar scattering, viscous gas drag from circumbinary disks and GW emission. Simulations that use subgrid dynamical friction instead of respositioning~\citep{2017MNRAS.470.1121T,2019MNRAS.482.2913B,2022MNRAS.513..670N}, find a substantial population of BHs that wander away from the local potential minima. Indeed, there is observational evidence of the presence of off-nuclear BHs in local dwarf galaxies~\citep{2020ApJ...888...36R}. This is overall important because if a substantial number of BHs wander far enough away from their original host galaxies, it will reduce the overall BH occupation fractions. Gravitational recoil can also lead to BHs getting kicked out of their hosts, further reducing the occupation fractions~\citep{2010MNRAS.404.2143V,2016MNRAS.456..961B,2020ApJ...896...72D}. Since we artificially suppress the possibility of BH wandering, we can interpret our predicted occupations fractions as upper limits for a given seed model.  

In the context of this work, the overall takeaway is the following: We clearly demonstrate that there is strong sensitivity to BH seeding within $\lesssim10^6~M_{\odot}$ local BH populations. However, whether we can disentangle these seeding signatures against signatures of other physics, particularly accretion, feedback and dynamics modeling, remains to be seen. We shall address this in future work. 

\section{Summary and Conclusions}
\label{Conclusions sec}

In this paper, we use the \texttt{BRAHMA} cosmological simulations to determine what can be inferred about BH seeding from observations of local BH populations. Our simulations are $[18~\rm Mpc]^3$ in volume and seed BHs close to the gas mass resolution, with an initial mass of $\seedmass=1.5\times10^5~M_{\odot}$. These seeds are formed using an array of prescriptions that encompass a broad range of seeding scenarios for heavy seeds as well as lower mass seeds. 

Our first four boxes are the same as in \cite{2024MNRAS.533.1907B} wherein the $1.5\times10^5~M_{\odot}$ BHs are directly formed as heavy seeds. In this paper, we refer to them as ``direct heavy seeds" or DHSs. To seed the DHSs, we incrementally stack several seeding criteria motivated from conditions postulated to be crucial for DCBH formation. In the first box~(\texttt{SM5}), we place DHSs in all halos that contain a minimum amount~($5~\seedmass$) of dense and metal poor gas. In the second box~(\texttt{SM5_LW10}), we additionally require the dense and metal poor gas to be illuminated by LW radiation~($>10~J_{21}$). In the third box~(\texttt{SM5_LW10_LOWSPIN}), we further require the seed forming halos to have gas spins less than the Toomre instability threshold. Finally, we have a fourth box~(\texttt{SM5_LW10_LOWSPIN_RICH}) wherein seed formation is further restricted to halos in rich environments with at least one neighboring halo of comparable or higher mass. The last criterion is motivated by the possibility that the neighboring halo will presumably merge and dynamically heat the seed forming halo to create regions where DCBHs can be born without a strong LW radiation. 

In addition to the four boxes that seed DHSs, we have a fifth box~(\texttt{ESD:STOCHASTIC}) wherein we use our recently developed stochastic seed model~\citep{2024MNRAS.529.3768B} to initialize the $1.5\times10^5~M_{\odot}$ BHs as descendants of lower mass $\sim10^3~M_{\odot}$ seeds. Since the initial $1.5\times10^5~M_{\odot}$ BHs are not meant to be the \textit{true seeds}, they are referred to as ``extrapolated seed descendants" or ESDs.  In this model, the ESDs are stochastically placed in galaxies with a broad distribution of masses and  preferentially living in rich environments. The seed model was calibrated against highest resolution simulations from \cite{2024MNRAS.531.4311B} that explicitly resolved these $\sim10^3~M_{\odot}$ seeds and traced their growth. 

Having five simulation boxes that differ only in their BH seed models enabled us to systematically study the impact of seeding on local BH and AGN populations, for the first time using cosmological hydrodynamic simulations. Our key findings are as follows:  
\begin{itemize}

\item The relative contribution to BH growth from mergers vs. gas accretion is a key factor that determines the persistence of seeding signatures in BH populations. In the high-z~($z\gtrsim5$) Universe, the seeding signatures can persist across all BHs at least up to $\sim10^8~M_{\odot}$, since the BH growth is merger dominated. However, at $z\sim0$, they tend to get largely erased for $\gtrsim10^7~M_{\odot}$ BHs due to accretion-driven BH growth. In this regime, all the seed models approach the observational measurements for the BHMFs and $M_*-M_{bh}$ relations. 

\item At the lowest mass end of $\sim10^5-10^6~M_{\odot}$ BHs, our seed models do leave strong signatures even in the local Universe. This is because of two reasons: 1) There is a substantial population of local $\sim10^5~M_{\odot}$ BHs that  are ungrown relics of seeds forming at high redshifts~($z\sim5-10$); 2) The growth of these BHs from $\sim10^5~M_{\odot}$ to $\sim10^6~M_{\odot}$ is primarily driven by mergers instead of gas accretion even at $z\sim0$. The predicted seed model variations in the number densities of $\gtrsim10^5~M_{\odot}$ range from $0.02-0.4~\mathrm{Mpc}^{-3}$. For $\gtrsim10^6~M_{\odot}$ BHs, it ranges from $0.01-0.05~\mathrm{Mpc}^{-3}$.

\item  For the ESD model that represents unresolved $\sim10^3~M_{\odot}$ seeds, the abundances of $\sim10^6~M_{\odot}$ BHs 
and $M_{bh}/M_*$ ratios for $\sim10^5-10^6~M_{\odot}$ BHs, are broadly consistent with available observational measurements. This supports $\sim10^3$ seeds~(resembling NSC seeds or extremely massive Pop III seeds) as viable origins for local SMBH populations. 

\item Amongst the DHS models, the most optimistic \texttt{SM5} model tends to mildly overpredict the abundances of $\sim10^6~M_{\odot}$ BHs 
and $M_{bh}/M_*$ ratios for $\sim10^5-10^6~M_{\odot}$ BHs compared to current  observations. The \texttt{SM5_LW10} and \texttt{SM5_LW10_LOWSPIN} models that additionally include the LW flux criterion and the gas spin criteria, are broadly consistent with both of these observed quantities. The predictions for the most restrictive \texttt{SM5_LW10_LOWSPIN_RICH} seed model are at the lower end of the observational measurements. Overall, these results imply that the heavy DCBH seed models can be viable origins for the local SMBH populations if their formation efficiencies are similar to what is implicitly assumed by our DHS models. Notably, these DHS models are much more plausible origins for local SMBHs compared to several previously proposed DCBH formation scenarios that require much higher LW fluxes~($\gtrsim1000~J_{21}$).

\item The occupation fractions of $\gtrsim10^5~M_{\odot}$ and $\gtrsim10^6~M_{\odot}$ BHs in $M_*\sim10^9~M_{\odot}$ galaxies range from $\sim40-100~\%$ and $\sim20-100~\%$ respectively. However, the current observational inferences have too large uncertainties to place strong constraints on our seed models. 

\item For constraining seed models using AGN LFs, we find that the local Universe is more promising than the high-z Universe~($z\sim5-10$). This is because the AGN LFs are sensitive to seed models only at extremely low luminosities of $L_{\rm 2-10~keV}\sim10^{39}~\mathrm{erg~s^{-1}}$. These luminosities may be reachable only in the local Universe with possible upcoming missions such as AXIS. However, these populations will be far from our reach in the high redshift Universe~($z\gtrsim5$).

Overall, our simulations affirm current expectations that in the local Universe, BHs with masses around $\sim10^5-10^6~M_{\odot}$ residing in dwarf galaxies are expected to exhibit strong signatures of seeding. But it may still be challenging to ascertain whether the bulk of our SMBH populations started as low mass seeds, or heavy seeds that form more efficiently than canonical DCBH scenarios. GW events from high-z mergers of $\gtrsim10^5~M_{\odot}$ BHs may be able to break this degeneracy as it may be a promising probe for the efficiency of heavy seed formation. However, signatures of BH accretion, feedback and dynamics may also be degenerate with BH seeding. We shall explore all this in future work. Nevertheless, our results strongly suggest that continued searches for local BHs between $\sim10^5-10^6~M_{\odot}$, can provide strong constraints for BH seeding that would be complementary to the high-z observations from JWST and future GW observations of merging BH binaries with LISA.

\end{itemize}
\section*{Acknowledgements}
This research was supported in part by grant NSF PHY-2309135 to the Kavli Institute for Theoretical Physics (KITP). AKB also acknowledges the organizers of the KITP workshop “Cosmic Origins: The First Billion Years”, during which some of this research was developed. LB acknowledges support from NSF awards AST-1909933 \& AST-2307171 and Cottrell Scholar Award \#27553 from the Research Corporation for Science Advancement.
PT acknowledges support from NSF-AST 2008490.
RW acknowledges funding of a Leibniz Junior Research Group (project number J131/2022).  LH acknowledges support by the Simons Collaboration on ``Learning the Universe''. 
\section*{Data availablity}
The underlying data used in this work shall be made available upon reasonable request to the corresponding author.

\bibliography{references}

\end{document}